\documentclass[lettersize,journal]{IEEEtran}
\usepackage{amsmath,amsfonts}
\usepackage{algorithmic}
\usepackage{algorithm}
\usepackage{array}
\usepackage[caption=false,font=normalsize,labelfont=sf,textfont=sf]{subfig}
\usepackage{textcomp}
\usepackage{stfloats}
\usepackage{url}
\usepackage{verbatim}
\usepackage{graphicx}
\usepackage{amsmath}
\usepackage{multicol}
\usepackage{cite}
\usepackage{tikz}
\usetikzlibrary{shapes,arrows,calc}
\usepackage[acronym,nonumberlist]{glossaries}
\newacronym{NCS}{NCS}{Networked Control Systems}
\newacronym{FSP}{FSP}{Filtered Smith Predictor}
\newacronym{SP}{SP}{Smith Predictor}
\newacronym{MIMO}{MIMO}{Multiple-Input Multiple-Output}
\newacronym{LMI}{LMI}{Linear Matrix Inequality}
\newacronym{LTI}{LTI}{Linear Time-Invariant}
\newacronym{SLTV}{SLTV}{Stochastic Linear Time-Variant}
\newacronym{3SFSP}{3SFSP}{Stochastic State-Space Filtered Smith Predictor}
\newacronym{MSS}{MSS}{Mean-Square Stability}
\newacronym{CACCS}{CACCS}{Cooperative Adaptive Cruise Control System}
\newacronym{iid}{IID}{Independent and Identically Distributed}
\usepackage{color,xcolor}

\usepackage{soul}
\setstcolor{red}

\usepackage{mathtools}

\tikzstyle{block} = [draw, rectangle, 
    minimum height=3em, minimum width=6em]
\tikzstyle{sum} = [draw, circle, node distance=1cm]
\tikzstyle{input} = [coordinate]
\tikzstyle{output} = [coordinate]
\tikzstyle{pinstyle} = [pin edge={to-,thin,black}]

\usepackage{amsthm}

\theoremstyle{definition}
\newtheorem{definition}{Definition}[section]

\begin{document}

\title{Predictor-Based Compensators for Networked Control Systems with Stochastic Delays and Sampling Intervals}

\author{Matheus Wagner \thanks{Matheus Wagner is with the Software/Hardware Integration Lab, Federal University of Santa Catarina, Florianópolis, Brazil}, Marcelo M. Morato \thanks{Marcelo M. Morato is with the  Automation and Systems Department, Federal University of Santa Catarina, Florianópolis, Brazil and the Univ. Grenoble Alpes, CNRS, Grenoble INP (Institute of
Engineering Univ. Grenoble Alpes), GIPSA-lab, 38000
Grenoble, France}, Antônio Augusto Fröhlich \thanks{Antônio Augusto Fröhlich is with the Software/Hardware Integration Lab, Federal University of Santa Catarina, Florianópolis, Brazil},  Julio E. Normey-Rico \thanks{Julio E. Normey-Rico is with the  Automation and Systems Department, Federal University of Santa Catarina, Florianópolis, Brazil}}

%\author{Matheus Wagner, Marcelo M. Morato, Antônio Augusto Fröhlich and Julio E. Normey-Rico}

%\markboth{IEEE Transacions on Systems, Man, and Cybernetics: Systems}%


\maketitle

\begin{abstract}

The stochastic nature of time delays and sampling intervals in Networked Control Systems poses significant challenges for controller synthesis and analysis, often leading to conservative designs and degraded performance. This work presents a modeling approach for Linear Multiple-Input Multiple-Output Networked Control Systems and introduces a compensation scheme based on the Filtered Smith Predictor to mitigate the adverse effects of stochastic time delays on closed-loop performance. The proposed scheme is evaluated through numerical simulations of a well-established Cooperative Adaptive Cruise Control system. Results demonstrate that the compensator achieves near-ideal average closed-loop performance and significantly reduces response variability compared to a traditional Filtered Smith Predictor. Notably, it yields a 45\% reduction in worst-case tracking error signal energy relative to an ideal baseline system with no time delays and constant sampling intervals.

\end{abstract}


\section{Introduction}

\gls{NCS} consist of spatially distributed components connected via communication networks \cite{baillieul2007control}, offering interoperability across heterogeneous devices and meeting the growing computational demands of modern control systems \cite{moyne2007emergence}.

However, \gls{NCS} introduce challenges in controller analysis and design. Traditional digital control assumes fixed sampling and actuation intervals with constant time delays \cite{franklin2002feedback}, whereas networks with shared communication media and modern processors often lead to variable, stochastic delays and sampling intervals \cite{liu2019survey}.

To address these issues, various modeling and design approaches have been proposed that account for the stochastic nature of delays, either as random multiples of a fixed sampling interval \cite{Pang2021, steinberger2022robust, bahreini2020robust, haghighi2020practical, antunes2021frequency} or as continuous random variables \cite{hu2023stability, hu2022stabilization, su2022h}. While discrete models require added delays and may degrade performance, continuous-delay models reduce conservatism but are typically limited to delays shorter than the sampling period \cite{hu2021discretization, sun2021analysis}.

Among the many design strategies \cite{bahreini2020robust, haghighi2021static, sun2021analysis, hu2022stabilization, haghighi2020practical, su2022h, lu2023necessary}, predictor-based compensators, particularly the \gls{SP} and its variants, stand out for their effectiveness in mitigating delay-induced performance loss \cite{steinberger2022robust, bonala2017delay, batista2018performance, mo2021hidden, normey2022control}.

Recent \gls{SP} adaptations for stochastic \gls{NCS} \cite{morato2021novel, steinberger2022robust, gamal2016delay, bonala2017delay, cuenca2010approach, batista2018performance, mo2021hidden, wu2018novel} either approximate delays as fixed constants \cite{gamal2016delay, bonala2017delay, steinberger2022robust} or use delay measurements to adjust predictions \cite{cuenca2010approach, batista2018performance, mo2021hidden}. Yet, fixed-delay models may not capture true system behavior, while measurement-based schemes cannot anticipate post-actuation delays—both leading to conservative designs to ensure robustness.

Given the necessity for design and analysis procedures that take into account the stochastic nature of \gls{NCS}, this work proposes a modeling procedure and a time delay compensation approach for \gls{NCS}, whose temporal properties are modeled as continuous random variables. Accordingly, our \textbf{main contributions} are:

\begin{enumerate}
    \item  A new modeling procedure is proposed for linear \gls{MIMO} \gls{NCS} in which time delays, sampling intervals and actuation intervals are modeled as continuous random variables, without the restriction for time delays to be smaller than the sampling intervals.
    \item A controller design procedure employing a stochastic time delay compensator based on the \gls{FSP} architecture, along with the stability conditions for the resulting closed-loop system and numerical simulations for validation of the proposed approach.
\end{enumerate}

\noindent \textbf{Paper Organization.} Sec. \ref{sec:system-model} describes the proposed \gls{NCS} modeling procedure. Sec. \ref{sec:compensator-design} describes the proposed compensator design procedure based on the \gls{FSP}. Sec. \ref{sec:stability-analysis} presents the stability test for the closed-loop system with the compensator presented in Sec. \ref{sec:compensator-design}. Sec. \ref{sec:numerical-example} provides a numerical example for the presented modeling and controller design procedures. Finally, Sec. \ref{sec:conclusion} presents the concluding remarks of this work.
 
\section{Preliminaries}\label{sec:system-model}

Considering a \gls{LTI} plant, this section presents a \gls{NCS} model taking into account that the sampling instant sequence may not be strictly periodic, but corrupted with sensing jitter, and that the random delays experienced by different state variables may not be equal. Sec. \ref{sec:system-model:tasks} describes the characterization of sensing, controller and actuation instant sequences in \gls{NCS}, while Sec. \ref{sec:system-model:discretization} describes the procedure to transform the continuous-time representation of a dynamical system into a stochastic discrete-time representation for the \gls{NCS}. Finally, Sec. \ref{sec:system-model:time-delays} presents the procedure to obtain a state-space representation of stochastic time delays.

\subsection{Temporal behavior of \gls{NCS}}\label{sec:system-model:tasks}

The implementation of an \gls{NCS} relies on a distributed real-time task set, primarily composed of sensing, control computation, and actuation tasks that communicate over a network. Sensing tasks acquire sensor measurements, the control task computes control signals based on these measurements, and actuation tasks apply the signals to the plant via actuators. Each task generates timed events that are crucial for analyzing the closed-loop system behavior, as detailed in the following paragraphs.

The sequence of measurement events is an ordered sequence of instants in which the signals from each one of the sensors are sampled, formally defined as: 

\begin{definition}[Sensing sequence]\label{modeling:definition:sensing-sequence}
    The sensing sequence for the $i$-th  sensor in a control system, denoted by $\{t^{i}_{k}\}^{\infty}_{k=0}$, is the sequence of time instants in which a measurement is acquired by sampling the signal produced by the sensor \footnote{the notation $\{x_{k}\}^{\infty}_{k=0}$ stands for a sequence of elements ${x_{1},x_{2}, ... }$ indexed by the variable $k$, assuming integer values from zero to infinity}. 
\end{definition}

Sensing would ideally be performed periodically, but, in practice, the sequence of sampling instants may contain some form of uncertainty. This uncertainty is referred to as sensing jitter and is defined as: 

\begin{definition}[Sensing jitter]\label{modeling:definition:sensing-jitter}
    Given a sampling period $T^s$, the expected sensing sequence for the $i$-th  sensor in a control system is given by $t^{i, expected}_{k} \coloneq kT^s$ and the sensing jitter, denoted by $\delta t^{s,i}_{k}$, is defined as $\delta t^{i}_{k} \coloneq t^{i, expected}_{k} - t^{i}_{k}$.
\end{definition}

The fact that sensing sequences from different sensors may not be synchronized is captured as a constant offset for the $i$-th  sensing sequence denoted by $o^i$. Given the aforementioned properties, the sensing sequence for the $i$-th  sensor in a \gls{NCS} is described as: 

\begin{equation}\label{model:sampling-sequence}
    t^{i}_{k} \coloneq o^i + kT^{s} + \delta^{s,i}_k.
\end{equation}

A precondition for the start of the computation of the $k$-th  control signal is that the $k$-th  messages $m^{s,i}_k$ containing the measurements acquired at the $k$-th  sensing sequence instants are available for the controller to consume. Those messages, though, are subject to computational and communication delays, defined as:

\begin{definition}[Sensing delay]
    The $k$-th  sensing delay associated with the $i$-th  sensor in a \gls{NCS}, denoted by $D^{s,i}_{k}$, is the time interval between the $k$-th  element of the sensing sequence, denoted by $t^{i}_{k}$, and the instant in which a message containing the acquired measurement is produced and sent to the controller.
\end{definition}

\begin{definition}[Sensing message delay]
    The $k$-th  sensing message delay associated with the $i$-th  sensor in a \gls{NCS}, denoted by $D^{s,c,i}_{k}$, is the time interval between the instant in which the message containing the acquired measurement is sent and the instant in which the message is available for consumption by the controller.
\end{definition}

Once the measurements are available for the controller task, the computation of the control signal implies an additional time delay defined as: 

\begin{definition}[Control signal computation delay]
    The delay in the computation of the $k$-th  control signal, denoted by $D^{c}_{k}$, is the time interval between the instant in which the last among the $k$-th  messages $m^{s,i}_k$ containing the measurements of each sensor becomes available for consumption by the controller and the instant in which a set of messages is produced containing the different components of the control signal.
\end{definition}

Furthermore, each message $m^{a,j}_k$ produced by the controller task is sent to the corresponding actuator, resulting in computational and communication delays defined as follows:

\begin{definition}[Actuation message delay]
    The $k$-th  actuation message delay associated with the $j$-th  actuator in a \gls{NCS}, denoted by $D^{c,a,j}_{k}$, is the time interval between the instant in which the control signal computation finishes and the instant in which the corresponding actuation message is available for consumption by the actuator.
\end{definition}

\begin{definition}[Actuation delay]
    The $k$-th  actuation delay for the $j$-th  actuator, denoted by $D^{a,j}_{k}$, is the time interval between the instant in which the message $m^j_k$ containing the $j$-th  component of the control signal is available for consumption and the instant in which the actuation command is issued to the actuator.
\end{definition}

Finally, the sequence of actuation events is the ordered sequence of instants in which each component of the control signal is applied to the plant, formally defined as:

\begin{definition}[Actuation sequence]
    The actuation sequence for the $j$-th  actuator in a control system, denoted by $\{a^{j}_{k}\}^{\infty}_{k=0}$, is the sequence of time instants in which a component of the control signal is applied to the plant through the actuator.
\end{definition}

The definition of the actuation sequence for the $j$-th  actuator follows directly from the instants in which the command associated with the $j$-th  component of the control signal is issued to the respective actuator; hence, such a sequence can be defined as: 

\begin{equation}\label{model:actuation-task-activation-time}
        a^{j}_{k} \coloneq \max_{i}\{o^i + kT^{s} + \delta^{s,i}_k + D^{s,i}_k + D^{s,c,i}_{k}\} + D^{c}_{k} + D^{c,a,j}_{k} + D^{a,j}_k.
\end{equation}

The actuation sequence is of major importance for modeling the dynamic behavior of a \gls{NCS}, as it encapsulates the effects of every component of the end-to-end time delay, from measurement acquisition to the issuance of commands to the actuator, defined as follows:

\begin{definition}[Time delay]\label{modeling:definition:time-delay}
    The time delay $D^{i}_k$ in a control system is defined by the time interval between the acquisition of a measurement and the first application of a control signal component to the plant, such that $D^{i}_k = \min_{j}\{a^{j}_{k}\}-t^{i}_k$.
\end{definition}

Along with the described model for the temporal behavior of the \gls{NCS}, the following specifications are assumed to be fulfilled:

\begin{enumerate}
 \item The sensing sequences $t^{i}_k$ must be strictly monotonically increasing, hence $t^{i}_k - t^{i}_{k-1} > 0$.
 \item The sequences $t^{i}_k$ must be loosely synchronized, meaning that given $t_k = \min_{i} \{t^i_{k}\}$ and  $t_{k+1} = \min_{i} \{t^i_{k+1}\}$, then $t_k \leq t^i_{k} < t_{k+1}$ for all $k$.
 \item The sequence of instants in which the control signal computation starts must be strictly monotonically increasing, hence $c_{k} - c_{k-1} > 0$.
 \item The actuation sequences $a^{j}_k$ must be strictly monotonically increasing, hence $a^{j}_k - a^{j}_{k-1} > 0$.
\item The actuation sequences $a^{j}_k$ must be loosely synchronized, meaning that given $a_k = \min_{j} \{a^j_{k}\}$ and  $a_{k+1} = \min_{j} \{a^j_{k+1}\}$, then $a_k \leq a^j_{k} < a_{k+1}$ for all $k$.
\end{enumerate}

The set of specifications proposed above is considered in order to simplify the analytical treatment of the model, but is also motivated by practical considerations. Specifications (1), (3) and (4) ensure that the ordering of control signal computation and actuation events matches the ordering of sampling events. Specifications (2) and (5) enforce that every control signal computed based on the measurements from the $k{\textit{-th}}$ set of sensing jobs is applied to the plant before any of the components of the next control signal are applied to the plant by the next set of actuation jobs. In both cases, the fulfillment of the specifications implies an emulation of the system behavior that would be observed if all tasks were executed sequentially in a single computing device.

\subsection{Discretization of \gls{LTI} system dynamics}\label{sec:system-model:discretization}

Consider a continuous-time \gls{LTI} system described in state-space form by a set of ordinary differential equations, in which $N_a$ is the number of actuators of the system:

\begin{equation}\label{model:continuous-system}
    \boldsymbol{\dot{x}}(t) = A^o \boldsymbol{x}(t) + \sum_{j=1}^{N_a}\boldsymbol{b}^{o,j} u^j(t).
\end{equation}

The exact solution for the system of differential equations from Eq. \eqref{model:continuous-system} between two consecutive instants of the sequence $a_k = \min_{j} \{a^j_{k}\}$ is given by Eq. \eqref{model:exact-solution}, in which $\Delta a_{k+1} = a_{k+1} - a_{k}$. %Note that Eq. \eqref{model:exact-solution} depends both on $u^j(a_{k-1})$ and $u^j(a_{k})$, since each  control signal component is updated at different time instant $a^{j}_k \in [ a_k, a_{k+1})$, capturing the effects of actuation jitter. 

\begin{equation}\label{model:exact-solution}
    \begin{aligned}
        &\boldsymbol{x}(a_{k+1}) = e^{A^o\Delta a_{k+1}}\boldsymbol{x}(a_k)  \\
        &+ \sum_{j=1}^{N_a} \Biggl[ 
        \int_{a_{k}}^{a_{k+1}} e^{A^o(a_{k+1}-\tau)}\boldsymbol{b}^{o,j} u^j(a_{k})d\tau \\
        &- \int_{a_{k}}^{a^j_{k}}e^{A^o(a_{k+1}-\tau)}\boldsymbol{b}^{o,j} (u^j(a_{k})-u^j(a_{k-1}))d\tau\Biggr].
    \end{aligned}
\end{equation}

Eq. \eqref{model:exact-solution} can be written in the compact form presented in Eq. \eqref{model:exact-solution-compact}, in which $B_k=\text{col}(\boldsymbol{\beta}^j_k)$, $B^J_k=\text{col}(\boldsymbol{\beta}^{J,j}_k)$ and $A_{k}$ are defined as $A_{k} = e^{A^{o}\Delta a_{k+1}}$. Furthermore $\boldsymbol{x}(a_{k})$ is replaced by $\boldsymbol{x}_k$ for a simpler notation, and the symbol $\Delta\boldsymbol{u}_k = \boldsymbol{u}_k - \boldsymbol{u}_{k-1}$ is introduced. 

\begin{equation}\label{model:exact-solution-compact}
    \begin{aligned}
        \boldsymbol{x}_{k+1} &= A_k\boldsymbol{x}_k + B_k\boldsymbol{u}_k +  B^J_k \Delta\boldsymbol{u}_k. \\
        \boldsymbol{\beta}_{k}^{j} &= \int_{a_{k}}^{a_{k+1}} e^{A^o(a_{k+1}-\tau)}\boldsymbol{b}^{o,j}d\tau. \\
        \boldsymbol{\beta}_{k}^{J,j} &= -\int_{a_{k}}^{a^j_{k}}e^{A^o(a_{k+1}-\tau)}\boldsymbol{b}^{o,j}d\tau.
    \end{aligned}
\end{equation}

\begin{comment}
\begin{equation}\label{model:input-matrix}
    B_{k} = \begin{bmatrix}
        \vert & \vert & &\vert & \vert  \\
        \boldsymbol{\beta}_{k}^{0} & \boldsymbol{\beta}_{k}^{0} & ... & \boldsymbol{\beta}_{k}^{N_a-1}  & \boldsymbol{\beta}_{k}^{N_a}   \\
        \vert & \vert & &\vert & \vert  \\
    \end{bmatrix}
\end{equation}

\begin{equation}\label{model:jitter-matrix}
    B^J_{k} = \begin{bmatrix}
        \vert & \vert & &\vert & \vert  \\
        \boldsymbol{\beta}_{k}^{J,0} & \boldsymbol{\beta}_{k}^{J,0} & ... & \boldsymbol{\beta}_{k}^{J,N_a-1}  & \boldsymbol{\beta}_{k}^{J,N_a}   \\
        \vert & \vert & &\vert & \vert  \\
    \end{bmatrix}
\end{equation}

\begin{equation}\label{model:dynamic-matrix}
    A_{k} = e^{A^{o}\Delta a_{k+1}}
\end{equation}
\end{comment}

Next, in order to obtain a proper state-space \gls{NCS} model, the models representing the actuation jitter, $J^u$,  and the dynamics of the plant, $G$, presented in equations \eqref{model:actuation-jitter-system} and \eqref{model:plant-system}, respectively, must be arranged in a series connection, leading to the model of the system in the absence of time delays. The term $B^{w}_k\boldsymbol{w}_k$ is included in the system defined by Eq. \eqref{model:plant-system} to represent possible exogenous disturbances.

\begin{align}
  \label{model:actuation-jitter-system}
  J^u :\quad
  &
    \begin{cases}
              \boldsymbol{x}^{J^u}_{k+1} \!\!\!\!\!\! &= \boldsymbol{u}_k \\
                \boldsymbol{u}^{J^u}_{k} \!\!\! &= -B^J_k \boldsymbol{x}^{J^u}_{k} + (B_k + B^J_k)\boldsymbol{u}_k .
    \end{cases}
  \\
  \label{model:plant-system}
  G :\quad
  &
    \begin{cases}
      \boldsymbol{x}_{k+1} \!\!\!\!\!\!  &= A_k\boldsymbol{x}_k + \boldsymbol{u}^{J^u}_{k} + B^{w}_k\boldsymbol{w}_k \\
      \boldsymbol{y}_{k} \!\!\! &= \boldsymbol{x}_k .
    \end{cases} 
\end{align}

\subsection{Stochastic delays}\label{sec:system-model:time-delays}

For deterministic digital control systems, time delays can be incorporated into state-space models as a minimal realization of the $z^{-d}$ transfer function, having $d$ as a time delay expressed as a multiple of the sampling period. Next, we derive an extension of such a model that accounts for time delays represented by a continuous random variable.

Next, in order to obtain a state-space representation of stochastic time delays, we first denote a given time delay observation as $D^{i}_k$, which is decomposed into an integer and a fractional part - $N^{i,d}_k$ and $d^i_{k}$, respectively. The latter are, by construction, stochastic variables and defined as follows:

%The integer part is itself a random variable defined by Eq. \eqref{model:interger-delay-interval-count}. which is the number of backward steps $n$ from the $k{\textit{-th}}$ element of the $a_k$ sequence to be taken until the time delay $D^{i}_k$ is contained in the interval $[a_{k-n},a_{k-n+1})$. 

\begin{equation}\label{model:interger-delay-interval-count}
    N^{i,d}_k \coloneq \{ n > 1 ; a_{k-n} \leq a_{k} -  D^{i}_k < a_{k-n+1} \}.
\end{equation}

%The fractional part of the delay, defined in Eq. \eqref{model:fractional-delay}, is given by time interval between $a_{k-N^d_k}$ and $a_{k}-D^i_k$. Note that the fractional part of the delay is expressed as a time-advancement, as it will be more convenient for the modeling steps presented later in this section.

%The fractional part of the delay, defined in Eq. \eqref{model:fractional-delay}, is given by time interval between $a_{k-N^d_k}$ and $a_{k}-D^i_k$, which is the portion of the delay that cannot be captured as a sum of previous actuation intervals. Note that the fractional part of the delay is actually expressed as a time-advancement, as it will be more convenient for the modeling steps presented later in this section.

\begin{equation}\label{model:fractional-delay}
    d^{i}_k \coloneq \sum_{n=1}^{N^{i,d}_k}\Delta a_{k-n+1} - D^{i}_k .
\end{equation}

Note that the fractional part of the delay in Eq. \eqref{model:fractional-delay} is expressed as a time-advancement, as it will be more convenient for the modeling steps presented later in this Section.

To determine a proper state-space model to represent the stochastic time delay operator, it is necessary to relate the $i{\textit{-th}}$ component of $\boldsymbol{x}(a_{k-N^d_k} + d^i_k)$ to $\boldsymbol{x}(a_{k-N^d_k})$, modeling the fractional part of the delay, and then relate $\boldsymbol{x}(a_{k-N^d_k})$ to $\boldsymbol{x}(a_{k})$, modeling the integer part of the delay.

The model for the fractional part of the delay is based on the exact solution of Eq. \eqref{model:continuous-system} considering two different scenarios for each component of the control signal, one in which $a_{k-N^{i,d}_k} + d^i_k \in [a_{k-N^{i,d}_k}, a^j_{k-N^{i,d}_k})$ and another in which $a_{k-N^{i,d}_k} + d^i_k \in [a^j_{k-N^{i,d}_k}, a_{k-N^{i,d}_k+1})$. Both scenarios are captured in Eqs. \eqref{model:exact-solution-fractional-delay-simplified}-\eqref{model:forced-fractional-delay-plus}, in which the random variable $p^{ij}_k$ equals 1 if $a_{k-N^{i,d}_k} + d^i_k \in [a_{k-N^{i,d}_k}, a^j_{k-N^{i,d}_k})$ and zero otherwise, $\Gamma^{i-}_{k} = \text{col}(\boldsymbol{\Gamma}_{k}^{ij-})$ and $\Gamma^{i+}_{k}=\text{col}(\boldsymbol{\Gamma}_{k}^{ij+})$.

\begin{equation}\label{model:exact-solution-fractional-delay-simplified}
    \begin{aligned}
        \boldsymbol{x}(a_{k-N^{i,d}_k} + d^i_{k}) &= e^{A^{o}d^{i}_{k}}\boldsymbol{x}(a_{k-N^{i,d}_k}) + \\ & \Gamma^{i-}_{k} \boldsymbol{u}(a_{k-N^{i,d}_k -1}) +     \Gamma^{i+}_{k} \boldsymbol{u}(a_{k-N^{i,d}_k}) .
    \end{aligned}
\end{equation}

\begin{equation}\label{model:forced-fractional-delay-minus}
    \begin{aligned}
        &\boldsymbol{\Gamma}_{k}^{ij-} = p^{ij}_k\int_{a_{k-N^{i,d}_k}}^{a_{k-N^{i,d}_k} + d^i_{k}}e^{A^o(a_{k-N^{i,d}_k}+d^{i}_k-\tau)}\boldsymbol{b}^{o,j} d\tau + \\ & (1-p^{ij}_k) \int_{a_{k-N^{i,d}_k}}^{a_{k-N^{i,d}_k} + s_{k-N^{i,d}_k}^j}e^{A^o(a_{k-N^{i,d}_k}+d^{i}_k-\tau)}\boldsymbol{b}^{o,j} d\tau.
    \end{aligned}
\end{equation}

\begin{equation}\label{model:forced-fractional-delay-plus}
    \boldsymbol{\Gamma}_{k}^{ij+} = (1-p^{ij}_k)\int_{a_{k-N^{i,d}_k} + s_{k-N^{i,d}_k}^j}^{a_{k-N^{i,d}_k}+d^{i}_k} e^{A^o(a_{k-N^{i,d}_k}+d^{i}_k-\tau)}\boldsymbol{b}^{o,j}d\tau.
\end{equation}

The model for the integer time delays, defined in Eq. \eqref{model:generic-delay-state-equation}, relies on the fact that there exists $N^d_{\text{max}}$ such that $N^{i,d}_k \leq N^d_{\text{max}}$ for all $k > 0$, since the probability distributions of all components of the time delays are assumed to have compact support. 

\begin{equation}\label{model:generic-delay-state-equation}
    \begin{aligned}
              \boldsymbol{x}^l_{k+1} = A^{l}\boldsymbol{x}^l_k + &B^{d}\boldsymbol{u}^l_{k}. \\
              A^{l} = \setlength{\arraycolsep}{0.05cm} \begin{bmatrix}
            \boldsymbol{0}_{1 \times N^d_{\text{max}}-1} && \boldsymbol{0}_{1\times 1}\\
            \boldsymbol{I}_{N^d_{\text{max}}-1\times N^d_{\text{max}}-1} && \boldsymbol{0}_{N^d_{\text{max}}-1\times 1} \\
            \end{bmatrix},     &B^{l} = \begin{bmatrix}
            1 \\
            \boldsymbol{0}_{N^d_{\text{max}}-1\times 1} \\
            \end{bmatrix}.
    \end{aligned}    
\end{equation}

Given an admissible initial condition, the state vector of the system described by Eq. \eqref{model:generic-delay-state-equation} contains the delayed observations of a scalar input $u^l_k$ up to $u^l_{k-N^d_{\text{max}}}$. For a given delay observation $N^{i,d}_{k}$, the corresponding delayed input $u^l_{k-N^{i,d}_{k}}$ can be recovered by selecting a matrix $C^{l, N^{i,d}_{k}}$ such that the equality $C^{l, N^{i,d}_{k}}\boldsymbol{\boldsymbol{x}^l_k} = u^l_{k-N^{i,d}_{k}}$ is satisfied. This same rationale can be extended for vector inputs by simply considering the parallel composition of multiple instances of the system described by Eq. \eqref{model:generic-delay-state-equation}.

The dynamics of the term $e^{A^{o}d^{i}_{k}}\boldsymbol{x}(a_{k-N^d_k})$ from Eq. \eqref{model:exact-solution-fractional-delay-simplified} is captured by the system described in Eq. \eqref{model:state-fractional-delay-model}, considering the input of the system to be $\boldsymbol{u}^{lx}_{k} = \boldsymbol{x}(a_{k}) = \boldsymbol{x}_k$. The matrices $A^{lx}$ and $B^{lx}$ are given by the dynamic and input matrices of the parallel composition of the system described by Eq. \eqref{model:generic-delay-state-equation}, respectively. The output matrix $C^{lx}$ is a stochastic matrix given by $C^{lx} = e^{A^{o}d^{i}_{k}}C^{lx, N^d_{k}}$, with $C^{lx, N^{i,d}_{k}}$ being the stochastic output matrix of the integer delay system.

\begin{equation}\label{model:state-fractional-delay-model}
  L^{x} : 
    \begin{cases}
      \boldsymbol{x}^{lx}_{k+1} \!\!\!\!\!\! &= A^{lx}\boldsymbol{x}^{lx}_k + B^{lx}\boldsymbol{u}^{lx}_{k} \\
      \boldsymbol{y}^{lx}_{k} \!\!\! &= C^{lx}_k \boldsymbol{x}^{lx}_k .
    \end{cases} 
\end{equation}

The same rationale can be applied to obtain the dynamics of the terms from Eq. \eqref{model:exact-solution-fractional-delay-simplified} related to $\boldsymbol{u}(a_{k-N^d_k -1})$ and $\boldsymbol{u}(a_{k-N^d_k})$, which generates:

\begin{equation}\label{model:input-fractional-delay-model}
  L^{u} : 
    \begin{cases}
      \boldsymbol{x}^{lu}_{k+1} \!\!\!\!\!\! &= A^{lx}\boldsymbol{x}^{lu}_k + B^{lu}\boldsymbol{u}^{lu}_{k} \\
      \boldsymbol{y}^{lu}_{k} \!\!\! &= C^{lx}_k \boldsymbol{x}^{lu}_k .
    \end{cases} 
\end{equation}

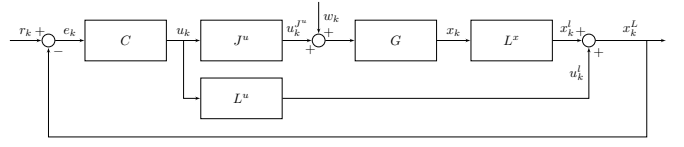
\begin{figure}
    \resizebox{\columnwidth}{!}{
        \begin{tikzpicture}[auto, node distance=2cm,>=latex']

    \node [input, name=reference] {};
    
    \node [sum, right of=reference] (trackingErrorSum) {};
    
    \node [block, right of=trackingErrorSum] (controller) {$C$};
    
    \node [block, right of=controller,
            node distance=3cm] (inputJitter) {$J^u$};
    
    \node [sum, right of=inputJitter,
            node distance=2cm] (disturbanceSum) {};

    \node [input, above of=disturbanceSum, node distance=1cm, name=disturbance] {};

    \node [block, below of=inputJitter,
            node distance=1.5cm] (inputDelay) {$L^u$};
    
    \node [block, right of=disturbanceSum,
            node distance=2cm] (plant) {$G$};

    \node [block, right of=plant,
            node distance=3cm] (stateDelay) {$L^x$};

    \node [sum, right of=stateDelay,
            node distance=2cm] (stateInputDelaySum) {};

    \node [output, right of=stateInputDelaySum] (output) {};

    \draw [draw,->] (reference) -- node {$r_k$} node[pos=0.95, anchor=south] {$+$} (trackingErrorSum);
    
    \draw [draw,->] (trackingErrorSum) -- node {$e_k$} (controller);

    \draw [draw,->] (controller) -- node [name=uc] {$u_k$} (inputJitter);

    \draw [draw,->] (uc) |- node {} (inputDelay);

    \draw [draw,->] (inputJitter) -- node {$u^{J^u}_k$} node[pos=0.95, anchor=north] {$+$}  (disturbanceSum);

    \draw [draw,->] (disturbance) -- node {$w_k$} node[pos=0.95, anchor=west] {$+$}  (disturbanceSum);

    \draw [draw,->] (disturbanceSum) -- node {} (plant);

    \draw [draw,->] (plant) -- node {$x_k$} (stateDelay);

    \draw [draw,->] (inputDelay) -| node [near end] {$u^{l}_k$}  node[pos=0.95, anchor=west] {$+$}  (stateInputDelaySum);

    \draw [draw,->] (stateDelay) -- node {$x^{l}_k$} node[pos=0.95, anchor=south] {$+$}  (stateInputDelaySum);

    \draw [draw,->] (stateInputDelaySum) -- node [name=xDelayed] {$x^{L}_k$} (output);

    \draw [draw,->] (16.5,0)--(16.5,-2.5)--(1,-2.5)--(1,-0.18) node[pos=0.95, anchor=west] {$-$};

    \end{tikzpicture}
    }   
    \caption{Closed loop system with fractional and integer time delays}
    \label{fig:stochastic-time-delay-block-diagram}
\end{figure}

By combining the control signal and state delay models presented, it is possible to define the undisturbed dynamics of a closed-loop system based solely on the state vector $\boldsymbol{x}_k$, the input vector $\boldsymbol{u}_k$ and their integer delayed values, while considering the stochastic nature of time delays. A block diagram of the resulting system representation is presented in Fig. \ref{fig:stochastic-time-delay-block-diagram}.

\section{Compensator design}\label{sec:compensator-design}

Sec. \ref{sec:system-model} introduces the notation and provides an overview of the proposed model for \gls{NCS}. This section presents the main contribution of this work: a design method for time-delay compensators for \gls{NCS} described by the proposed model.

\begin{figure}
    \centering
    % The block diagram code is probably more verbose than necessary
    \resizebox{0.85\columnwidth}{!}{
        \begin{tikzpicture}[auto, node distance=2cm,>=latex']
    
        \node [input, name=reference] {};

        \node [sum, right of=reference] (trackingErrorSum) {};

        \draw [draw,->] (reference) -- node {$r_k$} node[pos=0.95, anchor=south] {$+$} (trackingErrorSum);
        
        \node [block, right of=trackingErrorSum] (controller) {$C$};

        \draw [draw,->] (trackingErrorSum) -- node {$e_k$} (controller);
        
        \node [sum, right of=controller,
                node distance=2.5cm] (disturbanceSum) {};

        \draw [draw,->] (controller) -- node [name=uc]  {$u_k$} node[pos=0.90, anchor=south] {$+$} (disturbanceSum);
    
        \node [input, above of=disturbanceSum, node distance=1cm, name=disturbance] {};

        \draw [draw,->] (disturbance) -- node {$w_k$} node[pos=0.90, anchor=west] {$+$} (disturbanceSum);
        
        \node [block, right of=disturbanceSum,
                node distance=2cm] (plant) {$P$};

        \draw [draw,->] (disturbanceSum) -- (plant);

        \node [output, right of=plant, node distance=6cm] (output) {};
        
        \draw [draw,->] (plant) -- node {$x_k$} (output);
    
        \node [block, below of=plant,
                node distance=1.5cm] (plantModel) {$G_m$};

        \draw [draw,->] (uc) |- node {} (plantModel);

        \node [block, right of=plantModel,
                node distance=3cm] (stateDelay) {$L$};

        \draw [draw,->] (plantModel) -- node  [name=prediction] {$\bar{x}_k$} (stateDelay);

        \node [sum, below of=prediction,
                node distance=1.5cm] (stateEstimateSum) {};

        \node [sum, right of=stateDelay, node distance=2cm, name=predictionErrorSum] {};    

        \draw [draw,->] (stateDelay) -- node {} node[pos=0.80, anchor=south] {$-$} (predictionErrorSum);

        \draw [draw,->] (plant) -| node {} node[pos=0.95, anchor=west] {$+$} (predictionErrorSum);

        \draw [draw,->] (prediction) -- node {} node[pos=0.95, anchor=west] {$+$} (stateEstimateSum);
    
        \node [block, right of=stateEstimateSum,
                node distance=2cm] (filter) {$F$};

        \draw [draw,->] (predictionErrorSum) |- node {}(filter);    

        \draw [draw,->] (filter) -- node {} node[pos=0.95, anchor=north] {$+$} (stateEstimateSum);    

        \draw [draw,->] (stateEstimateSum) -| node {} node[pos=0.98, anchor=east] {$-$} (trackingErrorSum); 
        
        \end{tikzpicture}
    }   
    \caption{Filtered Smith predictor structure}
    \label{fig:fsp-block-diagram}
\end{figure}
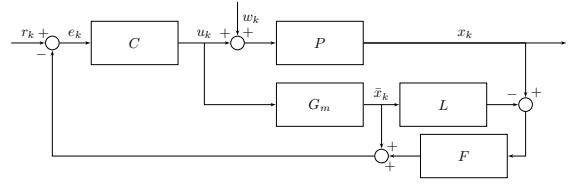

The original \gls{FSP} was designed to compensate for the effects of constant time delays in a closed-loop control system by using a predictor of the non-delayed outputs of the plant \cite{normey2007control}. It is composed of a primary controller $\boldsymbol{C}$, designed to stabilize the delay-free model of a \gls{LTI} plant $\boldsymbol{G_m}$, the actual plant $\boldsymbol{P}$, including time delays, a model for the plant time delay $\boldsymbol{L}$ and the prediction error filter $\boldsymbol{F}$ organized in the topology presented in Fig. \ref{fig:fsp-block-diagram}. 

The feedback signal {$\bar{x}_k$} supplied to the controller is given by the delay-free plant model, hence the system is able to anticipate the response of the time delay system and effectively track the reference signals as if there were no time delays. In reality, modeling errors and external disturbances are unavoidable. For this reason, the prediction error is used along with the delay-free model prediction to produce a more accurate estimate of the delay-free outputs. The role of the prediction error filter is to grant greater robustness to the overall system in exchange for slower response to disturbances. The procedure outlined in this section aims at deriving a variation of the \gls{FSP} that considers the state-space stochastic model for \gls{NCS} proposed in the previous sections, referred to as the \gls{3SFSP}.

The systematic design process of a \gls{FSP} consists of three parts. First, the primary controller is designed using any available technique, considering the delay-free model of the plant. Secondly, a model of the plant including time delays is established in order to build the predictor. Finally, the prediction error filter is tuned to meet a robustness specification and to avoid the appearance of undesired  plant model poles in the predictor dynamics, yielding a stable predictor even for unstable plant models.

\IEEEpubidadjcol

For the \gls{3SFSP}, the first step of the design process remains the same. For the modeling step, the models for the plant, input jitter, and time delay systems are replaced by the expected value of the stochastic model dynamics. With a slight abuse of notation, the expected value of the plant dynamics, $\bar{G} = \mathbf{E} \{ G \}$ is defined as in Eq. \eqref{compensator:expected-plant-system}. The expected value of the input jitter and time delay dynamics are defined in a similar manner, by taking the expected values of dynamic, input, output, and direct transfer matrices of each system, leading to the system structure presented in Fig. \ref{fig:3SFSP-block-diagram}. 

%For the \gls{3SFSP}, the first step of the design process remains the same. For the modeling step, since the models for the plant, input jitter and time delay systems are stochastic, they can not be employed for prediction directly and must be replaced by some deterministic model than can be evaluated in runtime. The deterministic system that achieves the minimum mean squared prediction error is the one given by the expected value of the stochastic model dynamics. With a slight abuse of notation, the expected value of the plant dynamics, $\bar{G} = \mathbf{E} \{ G \}$, is defined as in Eq. \eqref{compensator:expected-plant-system}. The expected value of the input jitter and time delay dynamics are defined in a similar manner, by taking the expected values of dynamic, input, output and direct transfer matrices of each system, leading to the system structure presented in Fig. \ref{fig:3SFSP-block-diagram}. 

\begin{equation}\label{compensator:expected-plant-system}
\bar{G} : 
    \begin{cases}
      \boldsymbol{\bar{x}}_{k+1} \!\!\!\!\!\! &= \mathbf{E}\{A_k\}\boldsymbol{\bar{x}}_k + \boldsymbol{\bar{u}}^{J^u}_{k} \\
      \boldsymbol{\bar{y}}_{k} \!\!\! &= \boldsymbol{\bar{x}}_k.
    \end{cases} 
\end{equation}

\begin{figure}
    \resizebox{\columnwidth}{!}{
        \begin{tikzpicture}[auto, node distance=2cm,>=latex']

    \node [input, name=reference] {};
    
    \node [sum, right of=reference] (trackingErrorSum) {};
    
    \node [block, right of=trackingErrorSum] (controller) {$C$};
    
    \node [block, right of=controller,
            node distance=3cm] (inputJitter) {$J^u$};
    
    \node [sum, right of=inputJitter,
            node distance=2cm] (disturbanceSum) {};

    \node [input, above of=disturbanceSum, node distance=1cm, name=disturbance] {};

    \node [block, below of=inputJitter,
            node distance=1.5cm] (inputDelay) {$L^u$};

    \node [block, below of=inputDelay,
            node distance=1.5cm] (averageInputDelay) {$\bar{L}^u$};

    \node [block, below of=averageInputDelay,
        node distance=1.5cm] (averageInputJitter) {$\bar{J}^u$};
    
    \node [block, right of=disturbanceSum,
            node distance=2cm] (plant) {$G$};

    \node [block, right of=plant,
            node distance=3cm] (stateDelay) {$L^x$};

    \node [sum, right of=stateDelay,
            node distance=2cm] (stateInputDelaySum) {};

    \node [block, right of=averageInputJitter,
            node distance=3cm] (plantModel) {$\bar{G}$};

    \node [block, right of=plantModel,
            node distance=4cm] (averageStateDelay) {$\bar{L}^x$};

    \node [sum, right of=averageStateDelay,
            node distance=2cm] (fractionalDelaySum) {};

    \node [block, below of=averageStateDelay,
            node distance=1.5cm] (filter) {$F$};

    \node [sum, left of=filter,
            node distance=2cm] (filterPredictorSum) {};

    \node [output, right of=stateInputDelaySum] (output) {};

    \draw [draw,->] (reference) -- node {$r_k$} node[pos=0.95, anchor=south] {$+$} (trackingErrorSum);
    
    \draw [draw,->] (trackingErrorSum) -- node {$e_k$} (controller);

    \draw [draw,->] (controller) -- node [name=uc] {$u_k$} (inputJitter);

    \draw [draw,->] (uc) |- node {} (inputDelay);

    \draw [draw,->] (uc) |- node {} (averageInputJitter);

    \draw [draw,->] (uc) |- node {} (averageInputDelay);

    \draw [draw,->] (inputJitter) -- node {$u^{J^u}_k$} node[pos=0.95, anchor=north] {$+$}  (disturbanceSum);

    \draw [draw,->] (disturbance) -- node {$w_k$} node[pos=0.95, anchor=west] {$+$}  (disturbanceSum);

    \draw [draw,->] (disturbanceSum) -- node {} (plant);

    \draw [draw,->] (plant) -- node {$x_k$} (stateDelay);

    \draw [draw,->] (inputDelay) -| node [near end] {$u^{l}_k$}  node[pos=0.95, anchor=west] {$+$}  (stateInputDelaySum);

    \draw [draw,->] (stateDelay) -- node {$x^{l}_k$} node[pos=0.95, anchor=south] {$+$}  (stateInputDelaySum);

    \draw [draw,->] (stateInputDelaySum) -- node [name=xDelayed] {$x^{L}_k$} (output);

    \draw [draw,->] (averageInputJitter) -- node {$\bar{u}^{\bar{J}^u}_k$} (plantModel);

    \draw [draw,->] (plantModel) -- node [name=xPredictor] {$\bar{x}_k$} (averageStateDelay);

    \draw [draw,->] (averageStateDelay) -- node {$\bar{x}^{l}_k$} (fractionalDelaySum);

    \draw [draw,->] (averageInputDelay) -| node {$\bar{u}^{l}_k$} (fractionalDelaySum);

    \node [sum, below of=xDelayed, node distance=4.825cm] (stateEstimateSum) {};

    \draw [draw,->] (xDelayed) -- node {} node[pos=0.99] {$-$}  (stateEstimateSum);

    \draw [draw,->] (fractionalDelaySum) -- node {$\bar{x}^{L}_k$} node[pos=0.80, anchor=north]  {$+$}  (stateEstimateSum);

    \draw [draw,->] (stateEstimateSum) |- node  {$\Delta\hat{x}_k$} (filter);

    \draw [draw,->] (xPredictor) -- node {} node[pos=0.88, anchor=east] {$+$}  (filterPredictorSum);

    \draw [draw,->] (filter) -- node {$\Delta\hat{x}^f_k$} node[pos=0.80, anchor=south] {$+$}  (filterPredictorSum);

    \draw [draw,->] (filterPredictorSum) -| node [near end] {$\hat{x}_k$} node[pos=0.99] {$-$}  (trackingErrorSum);
    
    \end{tikzpicture}
    }   
    \caption{\gls{3SFSP} structure}
    \label{fig:3SFSP-block-diagram}
\end{figure}

The filter design process is carried out similarly to the original \gls{FSP}, but considering the state-space model given by the expected value dynamics of the plant along with the actuation jitter and time delays. Consider the alternative structure for the predictor and time delay models presented in Fig.  \ref{fig:predictor-block-diagram}. The average plant, jitter, state delay, input delay and prediction error filter models can be described by the transfer matrices $\bar{G}(z)$, $\bar{J}^{u}(z)$, $\bar{L}^{x}(z)$, $\bar{L}^{u}(z)$ and $F(z)$, respectively.

\begin{comment}
    \begin{equation}\label{compensator:plant-transfer-matrix}
    \bar{G}(z) = \begin{bmatrix}
        \bar{G}_{11}(z) && ... && \bar{G}_{1N}(z) \\
        ... && ... && ... \\
        \bar{G}_{N1}(z) && ... && \bar{G}_{NN}(z)
    \end{bmatrix}
\end{equation}

\begin{equation}\label{compensator:jitter-transfer-matrix}
    \bar{J}^{u}(z) = \begin{bmatrix}
        \bar{J}^{u}_{11}(z) && ... && \bar{J}^{u}_{1Na}(z) \\
        ... && ... && ... \\
        \bar{J}^{u}_{N1}(z) && ... && \bar{J}^{u}_{NNa}(z)
    \end{bmatrix}
\end{equation}

\begin{equation}\label{compensator:state-delay-transfer-matrix}
    \bar{L}^{x}(z) = \begin{bmatrix}
        \bar{L}^{x}_{11}(z) && ... && \bar{L}^{x}_{1N}(z) \\
        ... && ... && ... \\
        \bar{L}^{x}_{N1}(z) && ... && \bar{L}^{x}_{NN}(z)
    \end{bmatrix}
\end{equation}

\begin{equation}\label{compensator:input-delay-transfer-matrix}
    \bar{L}^{u}(z) = \begin{bmatrix}
        \bar{L}^{u}_{11}(z) && ... && \bar{L}^{u}_{1N}(z) \\
        ... && ... && ... \\
        \bar{L}^{u}_{N1}(z) && ... && \bar{L}^{u}_{NN}(z)
    \end{bmatrix}
\end{equation}

\begin{equation}\label{compensator:filter-transfer-matrix}
    F(z) = \begin{bmatrix}
        F_{11}(z) && ... && F_{1N}(z) \\
        ... && ... && ... \\
        F_{N1}(z) && ... && F_{NN}(z)
    \end{bmatrix}
\end{equation}
\end{comment}

Since the transfer matrix $F(z)\bar{L}^u(z)$ represents a stable system, it will not take part in the rest of the analysis. The input-output behavior of the remainder of the predictor structure is given by the transfer matrix from Eq. \eqref{compensator:predictor-transfer-matrix}, in which $H(z) = \bar{G}(z)\bar{J}^{u}(z)$ and $P(z) = \bar{L}^{x}(z)H(z)$.

\begin{equation}\label{compensator:predictor-transfer-matrix}
    S(z) = H(z) - F(z)P(z).
\end{equation}

The transfer matrix $F$ must be designed to ensure that each element of the transfer matrix $S(z)$, as given by Eq. \eqref{compensator:predictor-transfer-matrix-elements}, is a stable transfer function.

\begin{equation}\label{compensator:predictor-transfer-matrix-elements}
    S_{nm}(z) = H_{nm}(z)-\sum_{r=1}^{N} F_{nr}(z)P_{rm}(z).
\end{equation}

The prediction model for the \gls{3SFSP} is obtained by the minimal realization of the S(z) transfer function, concluding the design.

\begin{figure}
    \centering
    % The block diagram code is probably more verbose than necessary
    \resizebox{0.9\columnwidth}{!}{
         \begin{tikzpicture}[auto, node distance=2cm,>=latex']

    \node [input, name=trackingError] {};
        
    \node [block, right of=trackingError] (controller) {$C$};
    
    \node [block, right of=controller,
            node distance=3cm] (averageInputJitter) {$\bar{J}^u$};
    
    \node [block, above of=averageInputJitter,
            node distance=1.5cm] (averageInputDelay) {$\bar{L}^u$};

    \node [block, right of=averageInputJitter,
            node distance=3cm] (plantModel) {$\bar{G}$};

    \node [block, right of=plantModel,
            node distance=3cm] (averageStateDelay) {$\bar{L}^x$};

    \node [block, right of=averageStateDelay,
            node distance=3cm] (stateFilter) {$F$};

    \node [block, right of=averageInputDelay,
            node distance=3cm] (inputFilter) {$F$};

    \node [sum, right of=stateFilter,
            node distance=2.2cm] (filterPredictorSum) {};

    \node [output, right of=filterPredictorSum, node distance=1.0cm] (output) {};

    \draw [draw,->] (trackingError) -- node {$e_k$} (controller);
    
    \draw [draw,->] (controller) -- node [name=uc, anchor=north] {$u_k$} (averageInputJitter);

    \draw [draw,->] (uc) |- node {} (averageInputDelay);

    \draw [draw,->] (averageInputJitter) -- node {$\bar{u}^{\bar{J}^u}_k$} (plantModel);

    \draw [draw,->] (plantModel) -- node [name=xPredictor] {$\bar{x}_k$} (averageStateDelay);

    \draw [draw,->] (averageStateDelay) -- node {$\bar{x}^{l}_k$} (stateFilter);

    \draw [draw,->] (averageInputDelay) -- node {$\bar{u}^{l}_k$} (inputFilter);

    \draw [draw,->] (inputFilter) -| node {$\bar{u}^{l,f}_k$} (filterPredictorSum)
     node[pos=0.95, anchor=east] {$-$};

    \draw [draw,->] (stateFilter) -- node {$\bar{x}^{l,f}_k$} (filterPredictorSum) node[pos=0.85, anchor=north] {$-$};

    \draw [draw,->] (9.5,0)--(9.5,-1)--(16.2,-1)--(16.2,-0.18) node[pos=0.95, anchor=west] {$+$};

    \draw [draw,->] (filterPredictorSum) -- node {$\bar{x}^{f}_k$} (output);

    \end{tikzpicture}   
    }
    \caption{Predictor alternative structure for filter design}
    \label{fig:predictor-block-diagram}
\end{figure}

\section{Stability test}\label{sec:stability-analysis}

There are several stability notions for stochastic dynamic systems, such as moment stability and almost sure stability \cite{nilsson1998real}. Of particular interest for the case of controller design is the notion of \gls{MSS}, since a stable system in \gls{MSS} is  guaranteed to have all of its observations to converge to a neighborhood of an equilibrium point. This section presents a simple procedure to prove the \gls{MSS} of a \gls{NCS} that employs the \gls{3SFSP} for control. 

Assuming that the reference signal has all of its components equal to zero, consider the closed-loop system described by the discrete-time \gls{SLTV}:

\begin{equation}\label{stability:closed-loop-dynamics}
    \boldsymbol{x}^{a}_{k+1} = A^{a}_k\boldsymbol{x}^{a}_{k}.
\end{equation}

The stability test to be derived provides the conditions in which $\lim_{k \to \infty} \mathrm{E}\bigl\{  \|\boldsymbol{x}^a_k\|^2 \bigl\} = 0$. Thus, consider the Lyapunov function candidate:

\begin{equation}\label{stability:lyapunov-function-candidate}
    V_{k} = \boldsymbol{x}^{aT}_{k} P \boldsymbol{x}^{a}_{k}.
\end{equation}

The expected dynamics of the Lyapunov function candidate can be expressed as:

\begin{equation}\label{stability:lyapunov-function-second-moment-dynamics}
    \mathrm{E}\bigl\{\Delta V_{k+1} | V_k\bigr\} = \boldsymbol{x}^{aT}_{k} \mathrm{E}\bigl\{ A^{aT}_k P A^{a}_k - P \bigr\} \boldsymbol{x}^{a}_{k}.
\end{equation}

A necessary condition for the system to be \gls{MSS} is that $\mathrm{E}\bigl\{\Delta V_{k+1} | V_k\bigr\} < 0$, which is equivalent to $\mathrm{E}\bigl\{ A^{aT}_k P A^{a}_k \bigr\} - P < 0$ \cite{sun2021analysis}. By vectorizing the inequality, it is possible to arrive at Eq. \eqref{stability:stability-condition-vectorized}, in which $\otimes$ stands for the Kronecker product and $\text{vec}(P)$ is the vectorization operator applied to $P$.

\begin{equation}\label{stability:stability-condition-vectorized}
        \mathrm{E}\bigl\{ A^{aT}_k \otimes A^{aT}_k \bigr\} \text{vec}(P) - \text{vec}(P) < \boldsymbol{0}.
\end{equation}

The term $\mathrm{E}\bigl\{ A^{aT}_k \otimes A^{aT}_k \bigr\}$ in Eq. \eqref{stability:stability-condition-vectorized} can be computed using Monte-Carlo methods and yields a deterministic matrix defined by $ \mathrm{E}\bigl\{ A^{aT}_k \otimes A^{aT}_k \bigr\} = \Xi^T \otimes \Xi$. By exploiting such definition, the stability condition can be rewritten as $\Xi^T P \Xi - P < 0$, implying that the system is stable as long as the spectral radius of $\Xi$ is smaller than one.

The direct evaluation of $\Xi$ may be computationally impractical, since the memory requirements of algorithms for computing the inverse Kronecker product are proportional to the fourth power of the dimension of $A^a$.

Notice that a key property of the Kronecker can be exploited to prove the stability of the system without the necessity of Kronecker product inversion. Given two matrices $A$ and $B$, whose eigenvalues are $\lambda_i$ for $i = 1,...,n$ and $\mu_j$ $j = 1,...,m$, respectively, the eigenvalues of $A \otimes B$ are $\lambda_i \mu_j$. Then, for any matrix $A$, the eigenvalues of $A \otimes A$ are given by $\lambda_i \lambda_j$. Hence, if the spectral radius of $\Xi^T \otimes \Xi$ is smaller than one, necessarily the spectral radius of $\Xi$ is also smaller than one.

In short, the stability condition described in Eq. \eqref{stability:closed-loop-dynamics} to be \gls{MSS} is that the spectral radius of $\mathrm{E}\bigl\{ A^{aT}_k \otimes A^{aT}_k \bigr\}$ is smaller than 1, and can be systematically evaluated through the following steps: 

\begin{enumerate}
    \item Given a set of samples from the stochastic processes that compose the \gls{NCS} model, compute the samples of $A^a_k$
    \item Compute the Monte-Carlo estimation of $\mathrm{E}\bigl\{ A^{aT}_k \otimes A^{aT}_k \bigr\}$ using the samples acquired in step 1.
    \item Compute the eigenvalues of $\mathrm{E}\bigl\{ A^{aT}_k \otimes A^{aT}_k \bigr\}$ and assert that the associated spectral radius is smaller than one.
\end{enumerate}

\section{Numerical example}\label{sec:numerical-example}

In order to demonstrate the effectiveness of the proposed approach, this section presents a numerical example considering the \gls{CACCS} from Ploeg et al. \cite{ploeg2011design}, commonly employed to evaluate \gls{NCS} design strategies \cite{saito2024network, li2017platoon}.

The objective of the \gls{CACCS} is to control the distance between vehicles, considering a time headway, through the cooperation of the vehicles in a platoon. For simplicity, this work considers only two vehicles for the numerical example. 

The continuous-time state-space model for the \gls{CACCS} is given by Eq. \eqref{example:continuous-dynamic-model} and Eq. \eqref{example:continuous-dynamic-model-parameters}, in which $\tau^{-1}$ is a parameter of the engine dynamics and $h$ is the time headway. The components of $\boldsymbol{x} = [x_1, x_2, x_3]$ represent the spacing, velocity, and acceleration differences between two vehicles, respectively, while the components of $\boldsymbol{u} = [u_1, u_2]$ represent the acceleration commands from the follower and the leader vehicles, respectively. %Note that the disturbance $\boldsymbol{w}$ is not present in the original model from \cite{ploeg2011design} and is included in this model so that the disturbance rejection capabilities of the proposed controller can be evaluated.

\begin{figure*}
\centering
\subfloat[]{\includegraphics[width=0.335\linewidth]{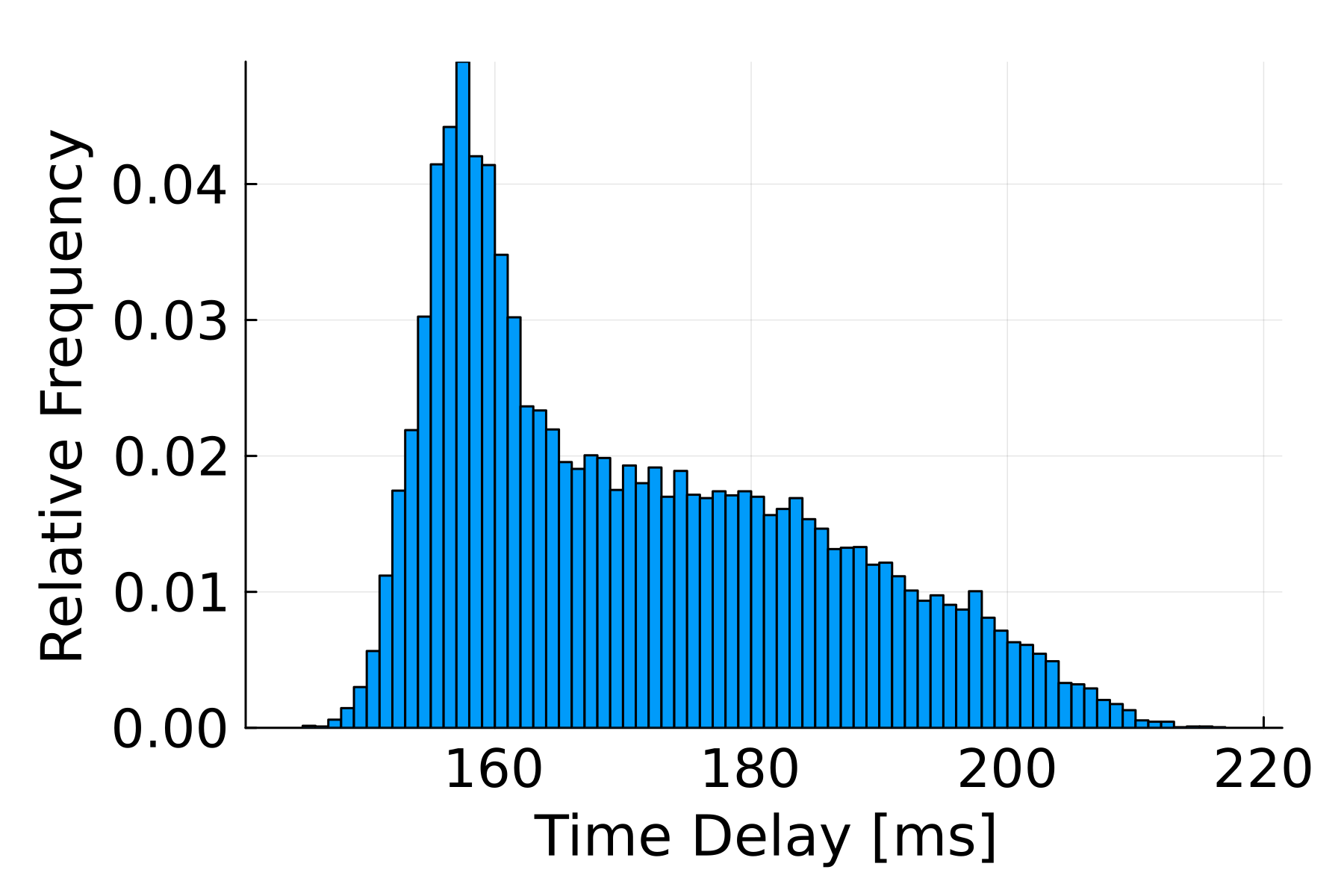}%
\label{fig:example:time-delay-distributions-e1}}
\subfloat[]{\includegraphics[width=0.335\linewidth]{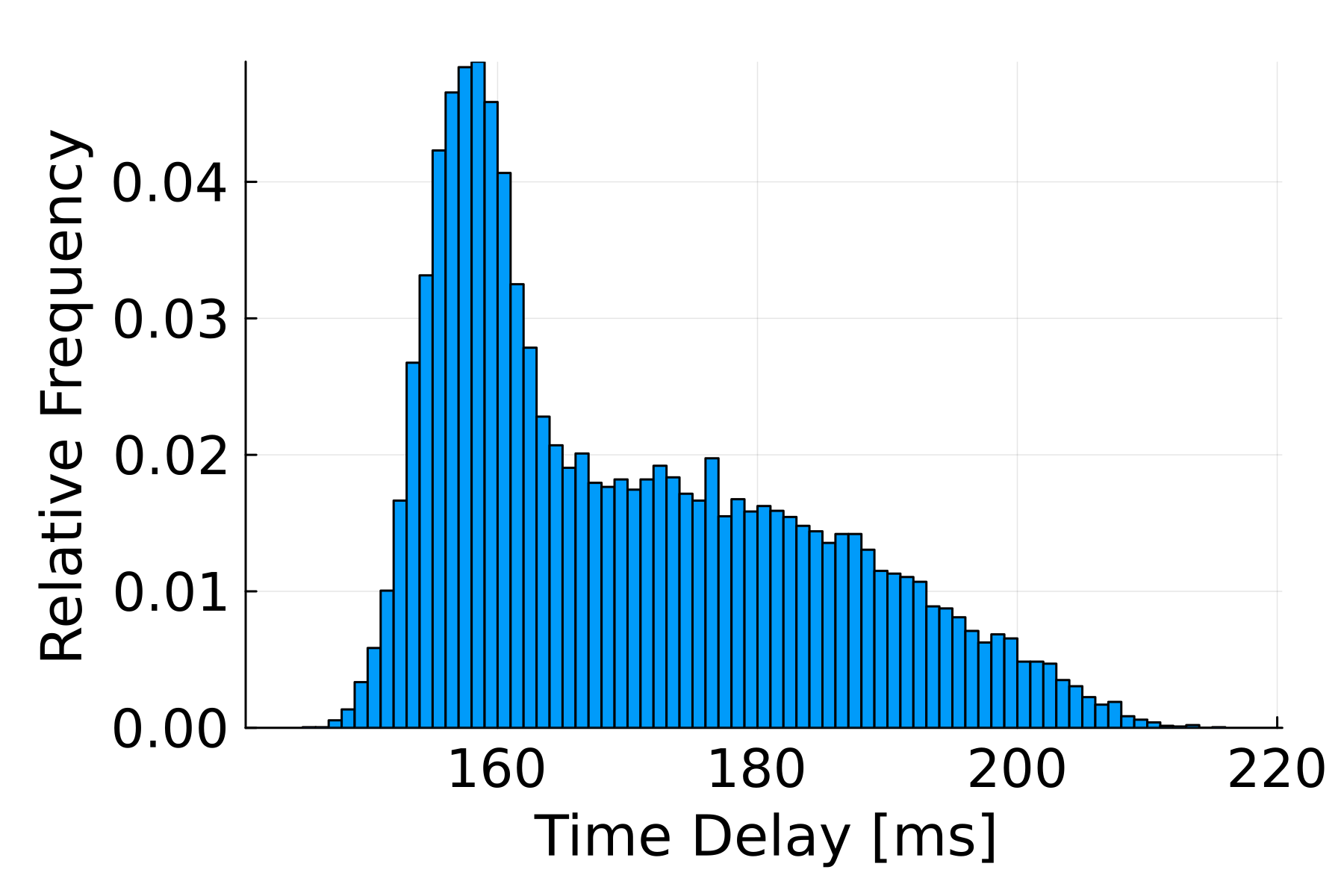}%
\label{fig:example:time-delay-distributions-e2}}
\subfloat[]{\includegraphics[width=0.335\linewidth]{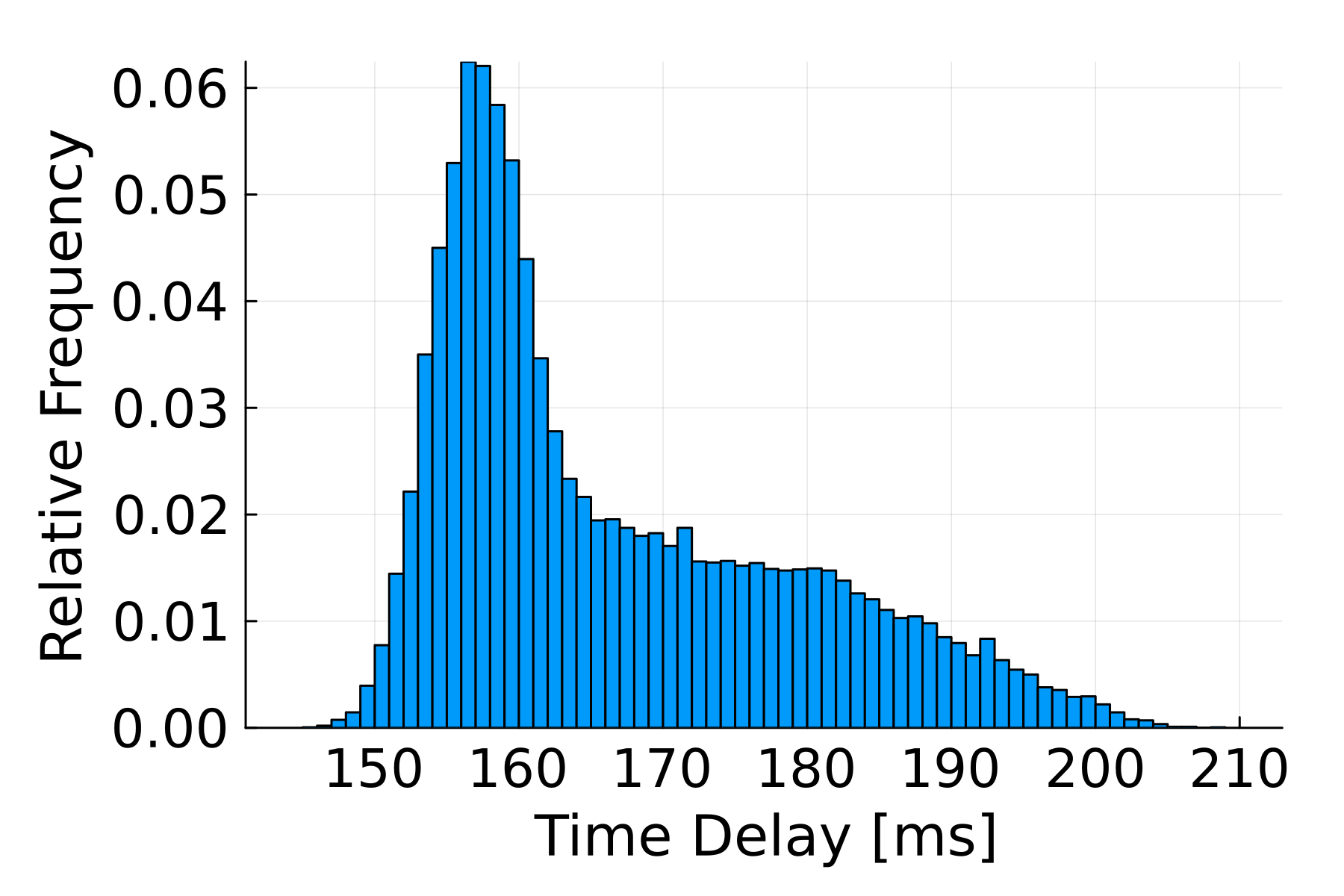}%
\label{fig:example:time-delay-distributions-e3}}
\caption{Time delay histograms for the numerical example. (a) Time delay observations for $x^1$. (b) Time delay observations for $x^2$. (c) Time delay observations for $x^3$}
\label{fig:example:time-delay-distributions}
\end{figure*}

\begin{figure}
\centering
\subfloat[]{\includegraphics[width=0.7\linewidth]{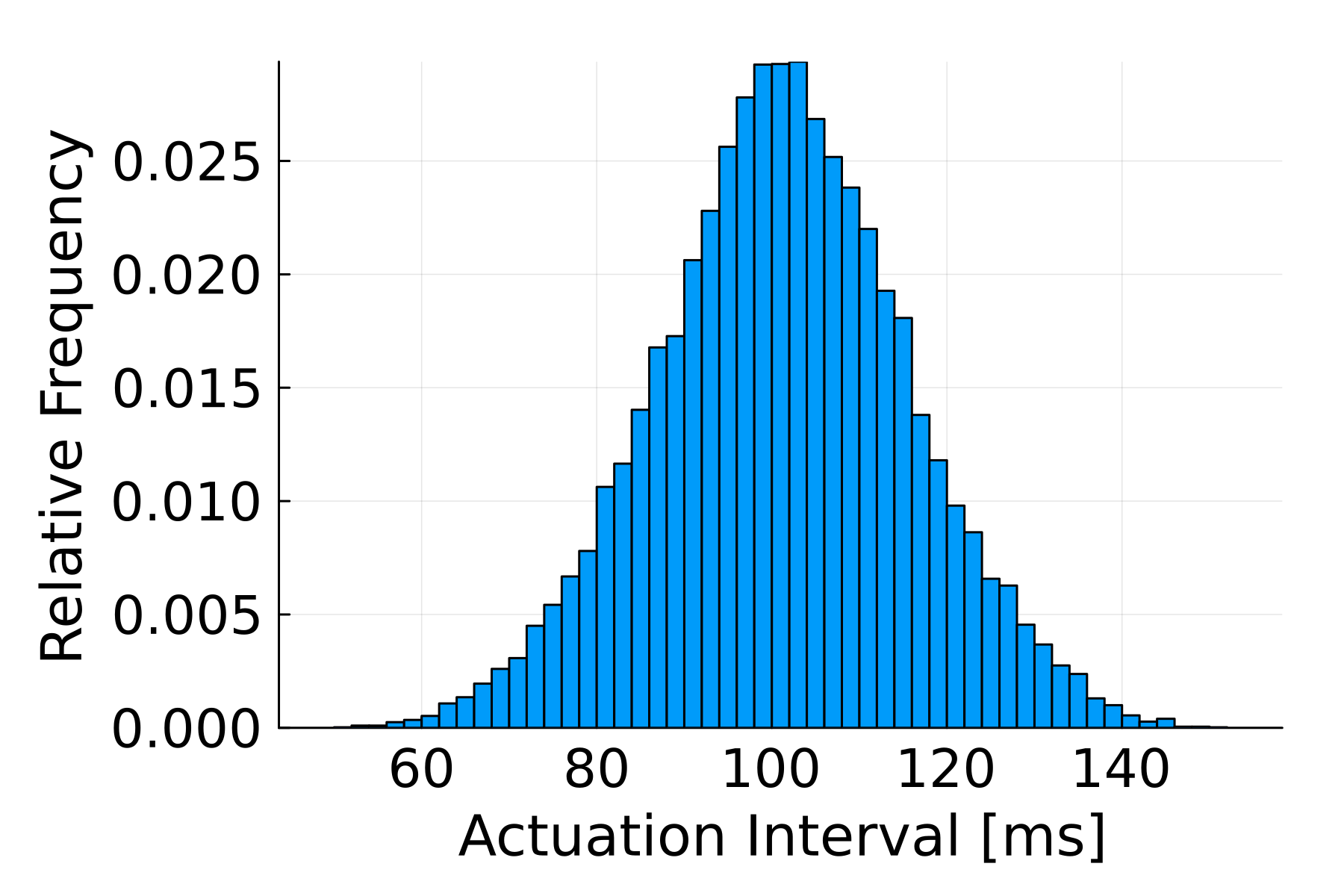}%
\label{fig:example:actuation-interval-distributions-u1}}
\hfil

\subfloat[]{\includegraphics[width=0.7\linewidth]{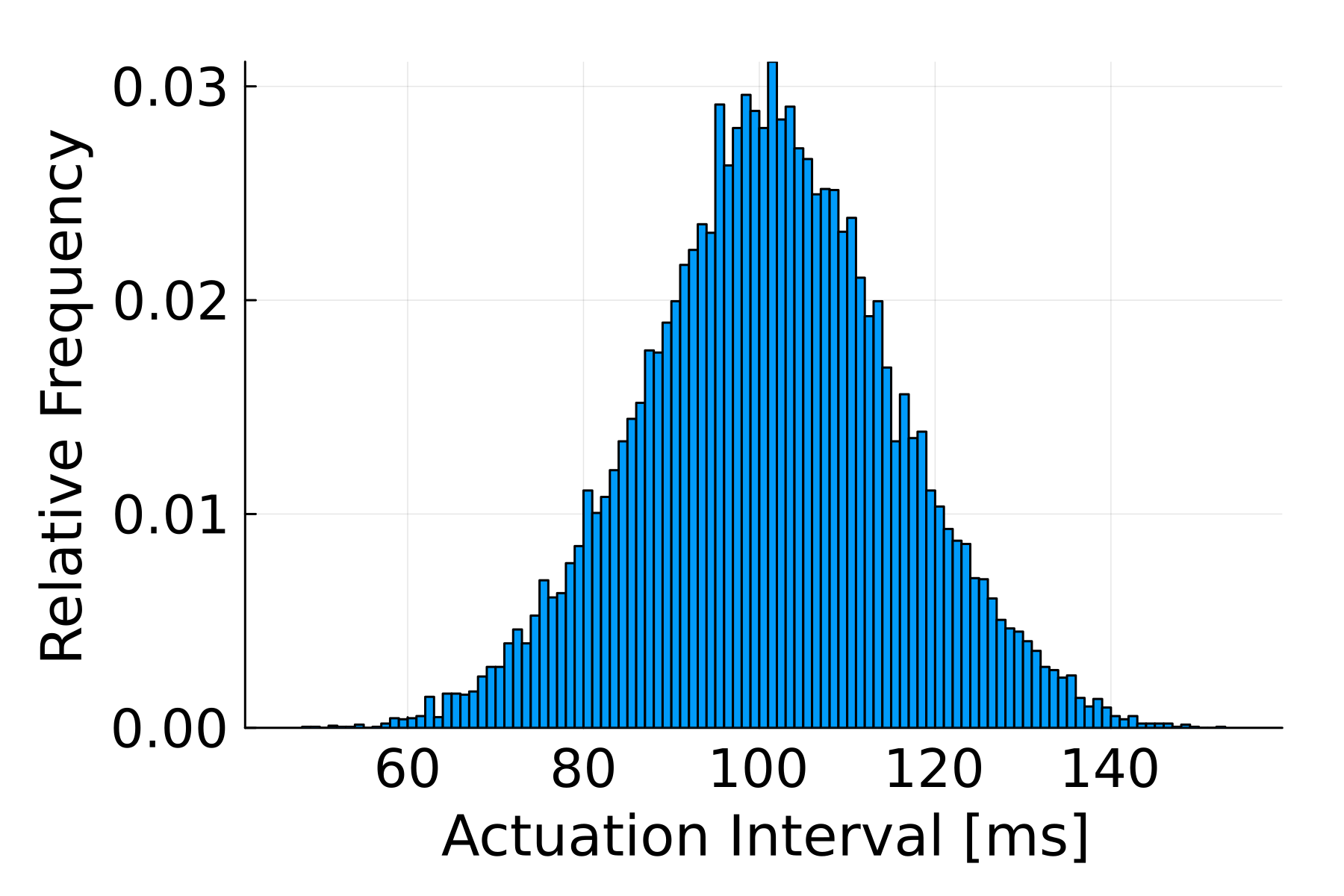}%
\label{fig:example:actuation-interval-distributions-u2}}
\caption{Actuation interval histograms for the numerical example. (a) Actuation interval observations for $u^1$. (b) Actuation interval observations for $u^2$.}
\label{fig:example:actuation-interval-distributions}
\end{figure}

\begin{equation}\label{example:continuous-dynamic-model}
    \dot{\boldsymbol{x}} = A^x \boldsymbol{x} + B^{x}\boldsymbol{u} + B^{w} \boldsymbol{w}.
\end{equation}

\begin{equation}\label{example:continuous-dynamic-model-parameters}
    \begin{aligned}
            &A^{x} = \begin{bmatrix}
            0 && 1 && 0 \\
            0 && 0 && 1 \\
            0 && 0 && -\tau^{-1}
        \end{bmatrix}, \\
            &B^{x} = \begin{bmatrix}
            0 && 0 \\
            0 && 0 \\
            h\tau^{-1} && -\tau^{-1}
        \end{bmatrix}, \quad
        B^{w} = \begin{bmatrix}
            0 && 0 \\
            0 && 0 \\
            1 && 1
        \end{bmatrix}.
    \end{aligned}
\end{equation}

For controller design, the continuous-time sis exactly discretized, considering a sampling period $T_s = 100$ ms. The primary controller structure consists of a dynamic controller, to include integral action, and a state-feedback gain designed using the discrete LQR procedure, resulting in the controller described in Eqs. \eqref{example:controller} and \eqref{example:controller-parameters}, in which $r_k$ is the inter-vehicle spacing reference to be tracked, hence $\boldsymbol{B}^r = [1,0,0]^T$. The matrices employed in the cost function for the LQR design were $Q = 2I/3$ and $R = I/3$.

\begin{equation}\label{example:controller}
    \begin{cases}
        \boldsymbol{x^c}_{k+1} \!\!\!\!\!\! &= A^c \boldsymbol{x^c}_{k}  + B^c (\boldsymbol{B}^r r_k -  \boldsymbol{x}_k) \\
        \boldsymbol{u}_k  &= -K^c\boldsymbol{x^c}_{k} + K^{x}(\boldsymbol{B}^r r_k - \boldsymbol{x}_k).
    \end{cases} 
\end{equation}

\begin{equation}\label{example:controller-parameters}
    \begin{aligned}
            A^{c} &= 1 ,\quad
            B^{c} = \begin{bmatrix}
            1  \\
            0  \\
            0
        \end{bmatrix} ,
        K^c = \begin{bmatrix}
           0.5986 \\ -0.8551
        \end{bmatrix} \quad\\
        K^{x} &= \begin{bmatrix}
            -7.8336 && -4.7042 && -1.1818 \\
            11.1908 && 6.7202 && 1.6883
        \end{bmatrix}.
    \end{aligned}
\end{equation}

The probability distributions for the temporal properties of the \gls{NCS} are presented in Table \ref{tab:example:ncs-parameters}. The Inverse Gaussian distribution was selected to characterize the execution time of tasks and transmission time of messages as previous works in the literature have pointed to it as a good fit for modeling response times \cite{zagalo2022response}. The truncated Inverse Gaussian distribution is a variation of the regular Inverse Gaussian distribution that has compact support. Fig. \ref{fig:example:time-delay-distributions} and Fig. \ref{fig:example:actuation-interval-distributions} present the histograms of the resulting time delay and actuation interval observations. Given the probability distributions for the \gls{NCS} temporal properties, the primary controller and the continuous-time model of the plant, the discrete-time stochastic models for the plant, input jitter and time delays were obtained using the procedure outlined in Sec. \ref{sec:system-model}.

\renewcommand{\arraystretch}{1.5}
\begin{table}[!t]
\caption{\gls{NCS} probability distribution parameters for the example}
\label{tab:example:ncs-parameters}
\begin{center}
    \begin{tabular}{|c|m{5cm}|}
    \hline
     \textbf{System Property} & \textbf{Distribution and Parameters} \\
     \hline
     Sensing Offset & Uniform (min=0s, max=20ms) \\
     \hline
     Sensing Jitter & Uniform (min=0s, max=50ms) \\
     \hline
     Sensing Task Delay & Truncated Inverse Gaussian ($min=14ms$, $max=17ms$, $\mu=16ms$, $\sigma = 0.85ms$) \\
     \hline
     Sensing Message Delay & Truncated Inverse Gaussian ($min=28ms$, $max=34ms$, $\mu=32ms$, $\sigma = 1.7ms$) \\
     \hline
     Control Computation  Delay & Truncated Inverse Gaussian ($min=42ms$, $max=51ms$, $\mu=48ms$, $\sigma = 2.5ms$) \\
     \hline
     Actuation Message Delay & Truncated Inverse Gaussian ($min=42ms$, $max=51ms$, $\mu=48ms$, $\sigma = 2.5ms$) \\
     \hline
     Actuation Task Delay & Truncated Inverse Gaussian ($min=14ms$, $max=17ms$, $\mu=16ms$, $\sigma = 0.85ms$) \\
     \hline
\end{tabular}
\end{center}

\end{table}

\begin{figure}
\centering
\subfloat[]{\includegraphics[width=0.7\linewidth]{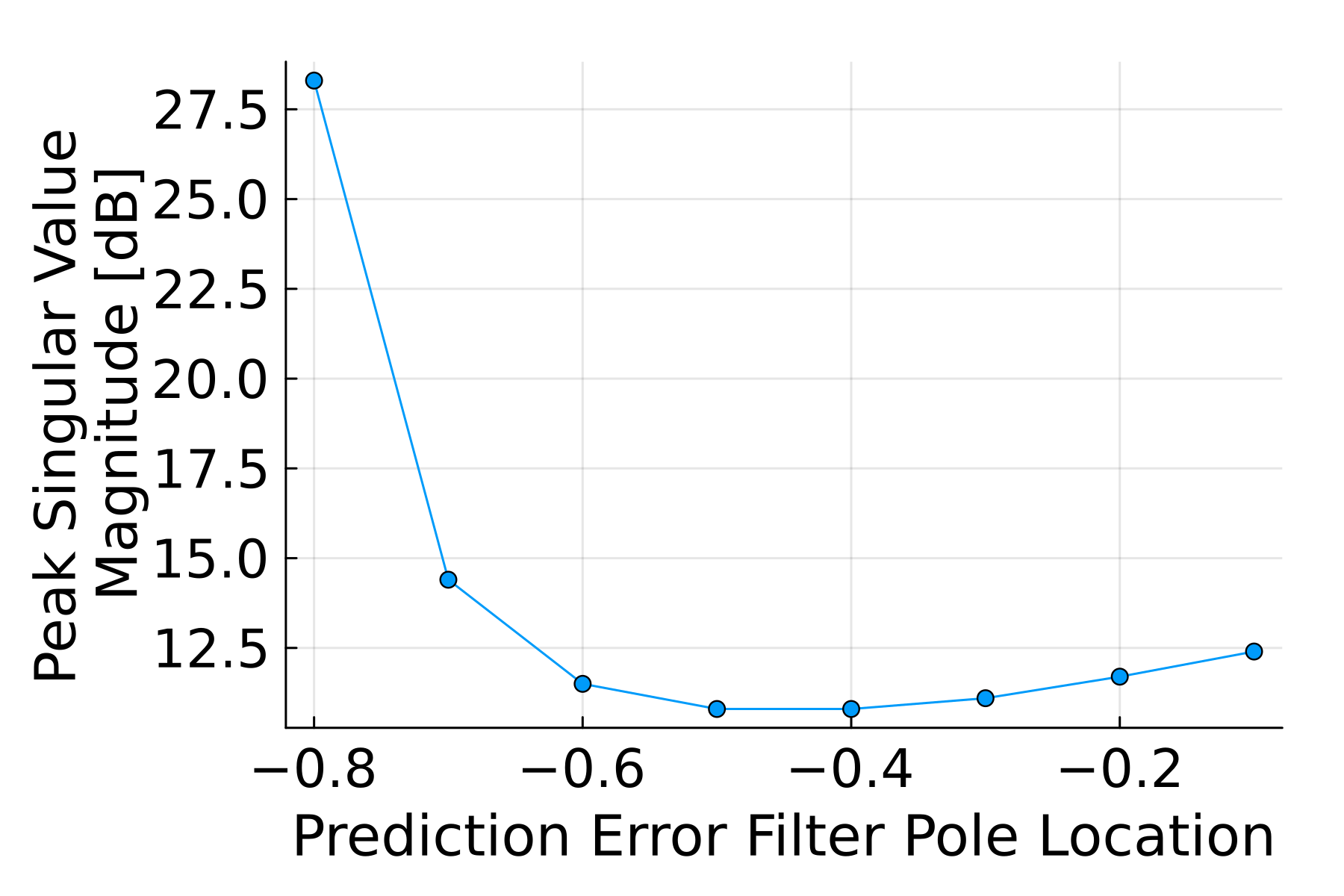}%
\label{fig:example:filter-parameteric-singular-value}}
\hfil

\subfloat[]{\includegraphics[width=0.7\linewidth]{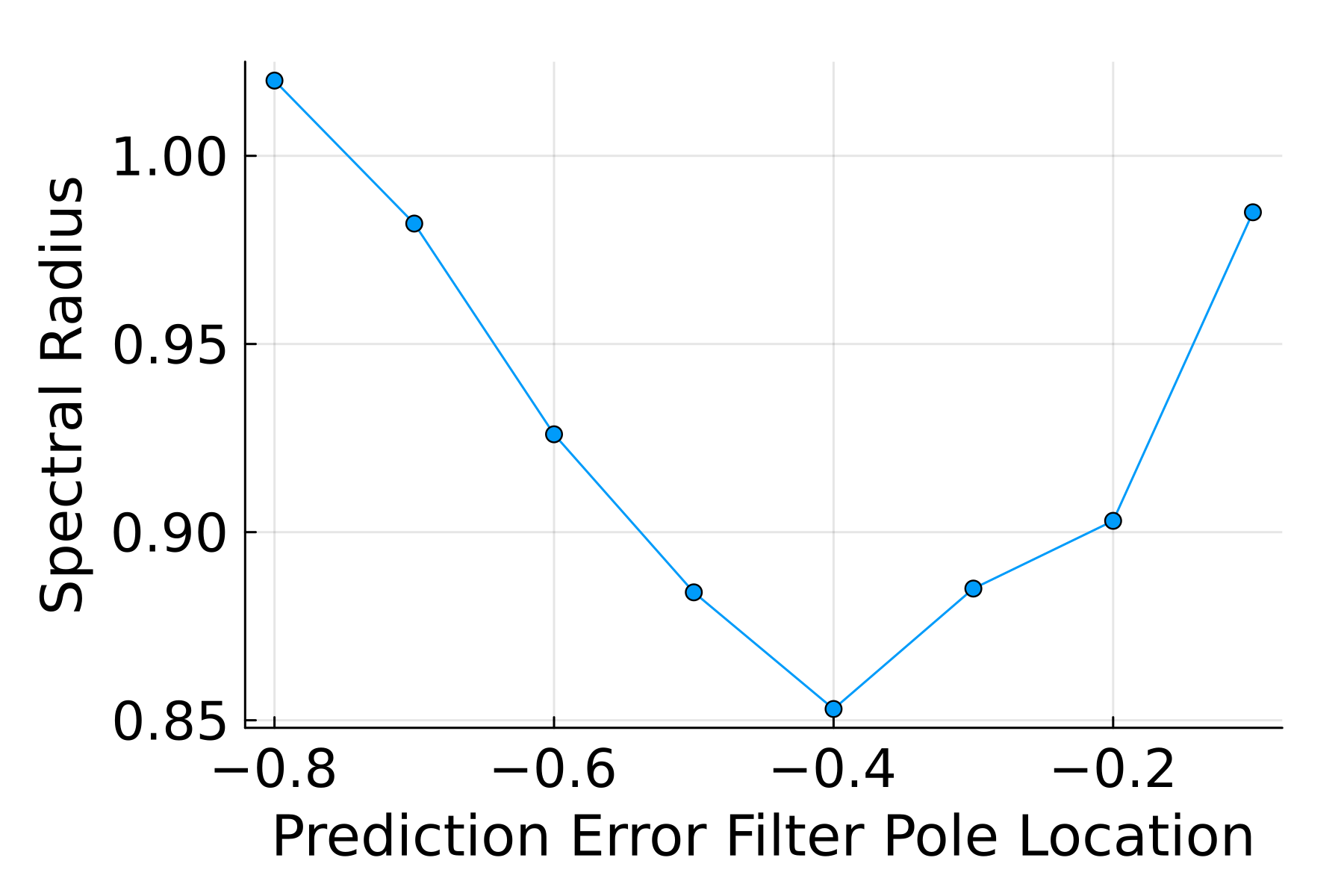}%
\label{fig:example:filter-parameteric-spectral-radius}}
\caption{Parametric analysis for selection of the prediction filter pole location. (a) Peak of the singular value frequency response from the input disturbance to output transfer function. (b) Spectral radius of the matrix $\Xi^T \otimes \Xi$.}
\label{fig:example:filter-parameteric-analysis}
\end{figure}

\begin{figure}
    \centering
    \includegraphics[width=0.65\linewidth]{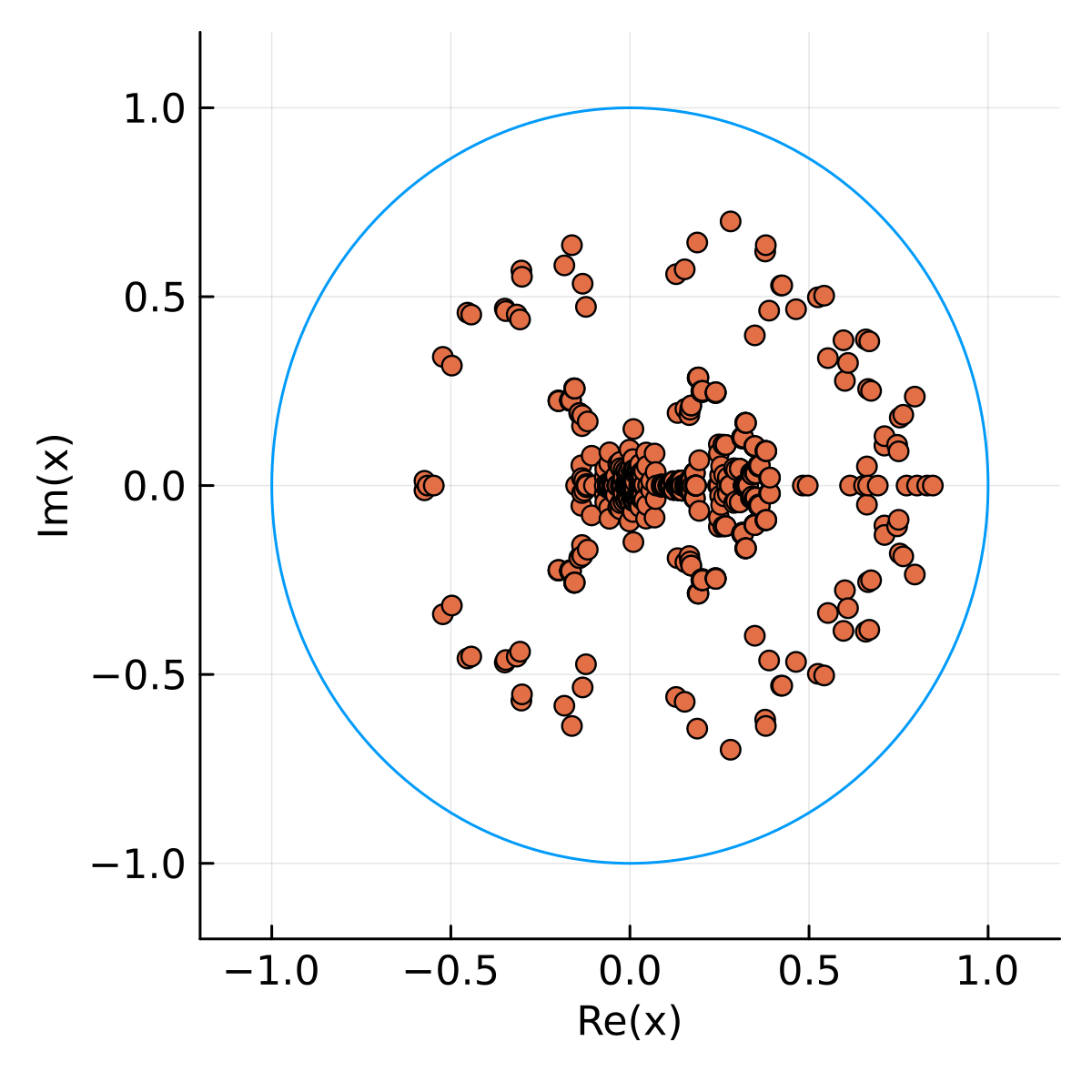}
    \caption{Pole diagram related to $\mathrm{E}\bigl\{ A^{aT}_k \otimes A^{aT}_k \bigr\}$ for the system of the numerical example.}
    \label{fig:closed-loop-poles}
\end{figure}

The filter design is carried out considering the diagonal filter structure from Eq. \eqref{example:filter-structure}. The zeros of the filter are designed to prevent the plant poles at $z=-1$, as well as the pole of the plant at $z = -0.819$, of appearing in the predictor dynamics. The poles of the filter are chosen based on a parametric analysis considering the spectral radius of the matrix $\Xi^T \otimes \Xi$, which is correlated with the set-point response performance, and the peak of the singular value frequency response from the expected value of the input disturbance to output transfer function, which is a metric for the robustness of the closed-loop system. As presented in Fig.  \ref{fig:example:filter-parameteric-analysis}, the parametric analysis results indicate that the best compromise between disturbance rejection and set-point responses is achieved when the pole location is given by $\alpha = 0.4$, which is the pole location selected for this design step.

\begin{equation}\label{example:filter-structure}
\begin{aligned}
        F_{11}(z) &= \frac{b_{14} z^4 + b_{13} z^3 + b_{12} z^2 + b_{11} z + b_{10}}{(z-\alpha)^5}. \\
        F_{22}(z) &= \frac{b_{14} z^4 + b_{13} z^3 + b_{12} z^2 + b_{11} z + b_{10}}{(z-\alpha)^5}. \\
        F_{33}(z) &= \frac{b_{31} z + b_{30}}{(z-\alpha)^2} .
\end{aligned}
\end{equation}

The stability test of the closed-loop system is carried out through a Monte-Carlo estimation using ten thousand observations of $A^{aT}_k$, which was shown to be enough for the convergence of the estimate. The resulting eigenvalues of $\mathrm{E}\bigl\{ A^{aT}_k \otimes A^{aT}_k \bigr\}$ are presented in Fig.  \ref{fig:closed-loop-poles}. By inspection, it is possible to conclude that stability is ensured according to the proposed test detailed in Sec. \ref{sec:stability-analysis}, as all poles lie within the unit circle and the spectral radius of $\mathrm{E}\bigl\{ A^{aT}_k \otimes A^{aT}_k \bigr\}$ is $0.856$.

\begin{figure*}
    \centering

    \subfloat[]{\includegraphics[width=0.335\linewidth]{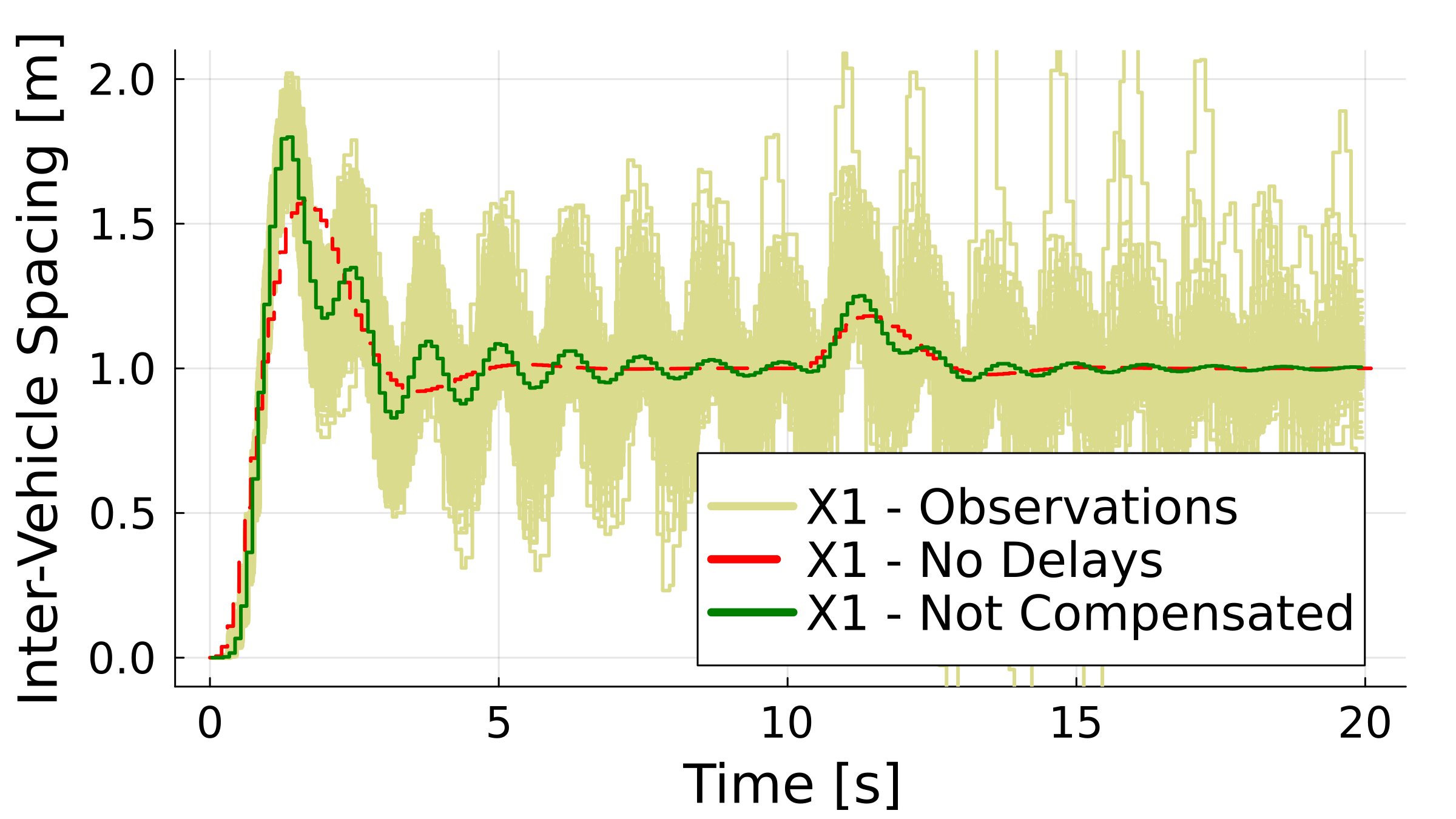}
    \label{fig:example:position-nc}}
    \subfloat[]{\includegraphics[width=0.335\linewidth]{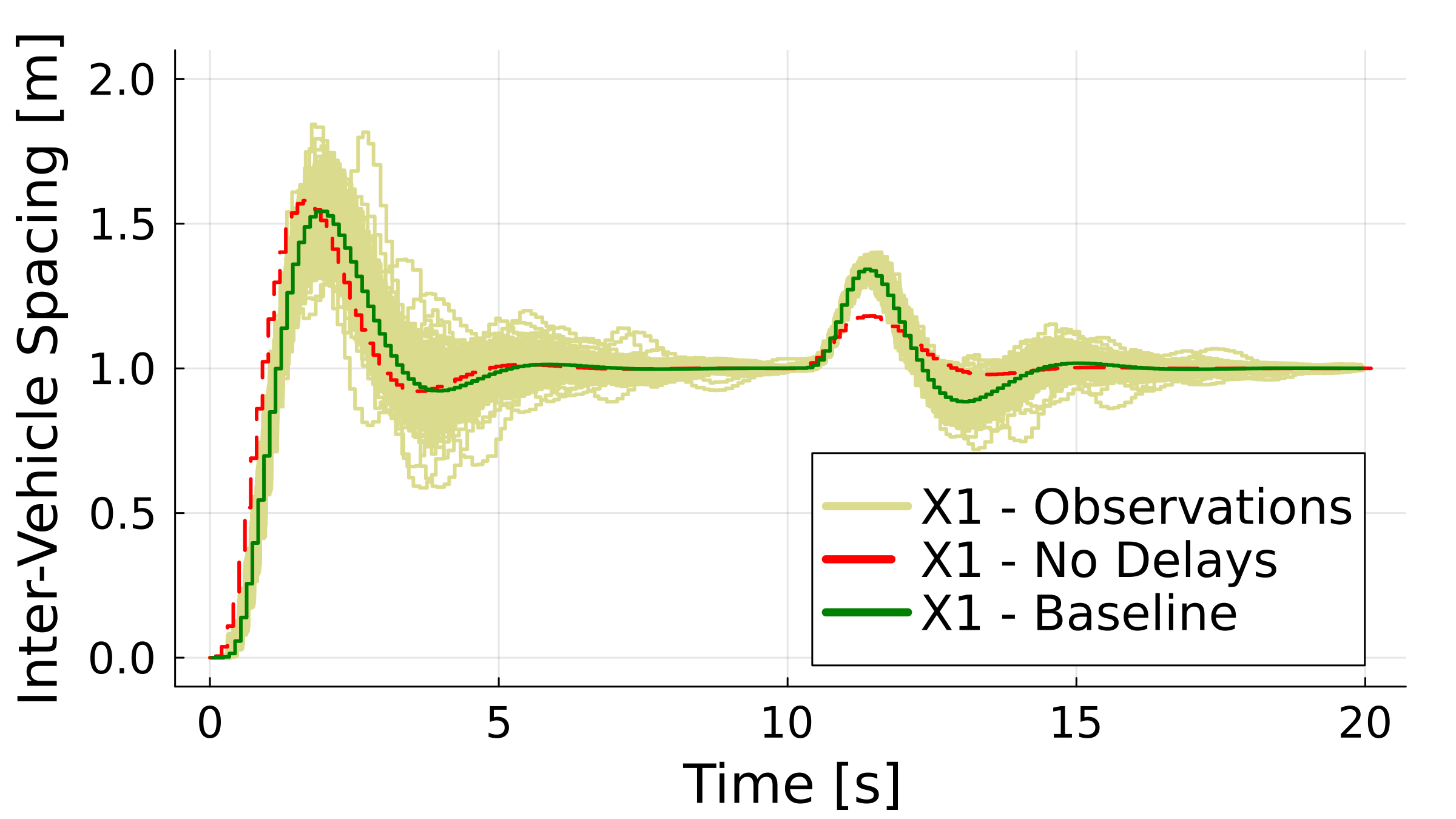}
    \label{fig:example:position-base}}
    \subfloat[]{\includegraphics[width=0.335\linewidth]{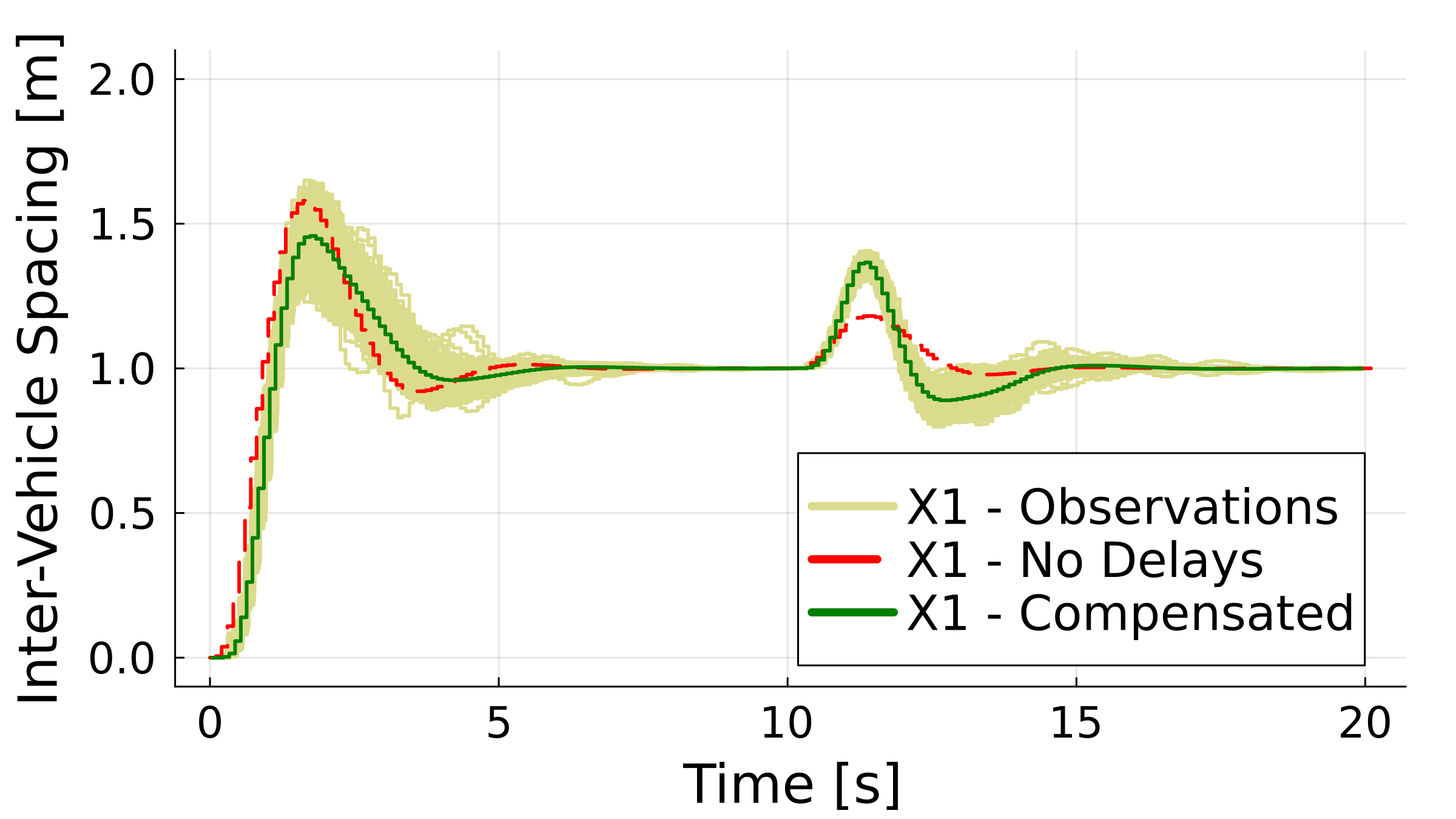}
    \label{fig:example:position-c}}

    \subfloat[]{\includegraphics[width=0.335\linewidth]{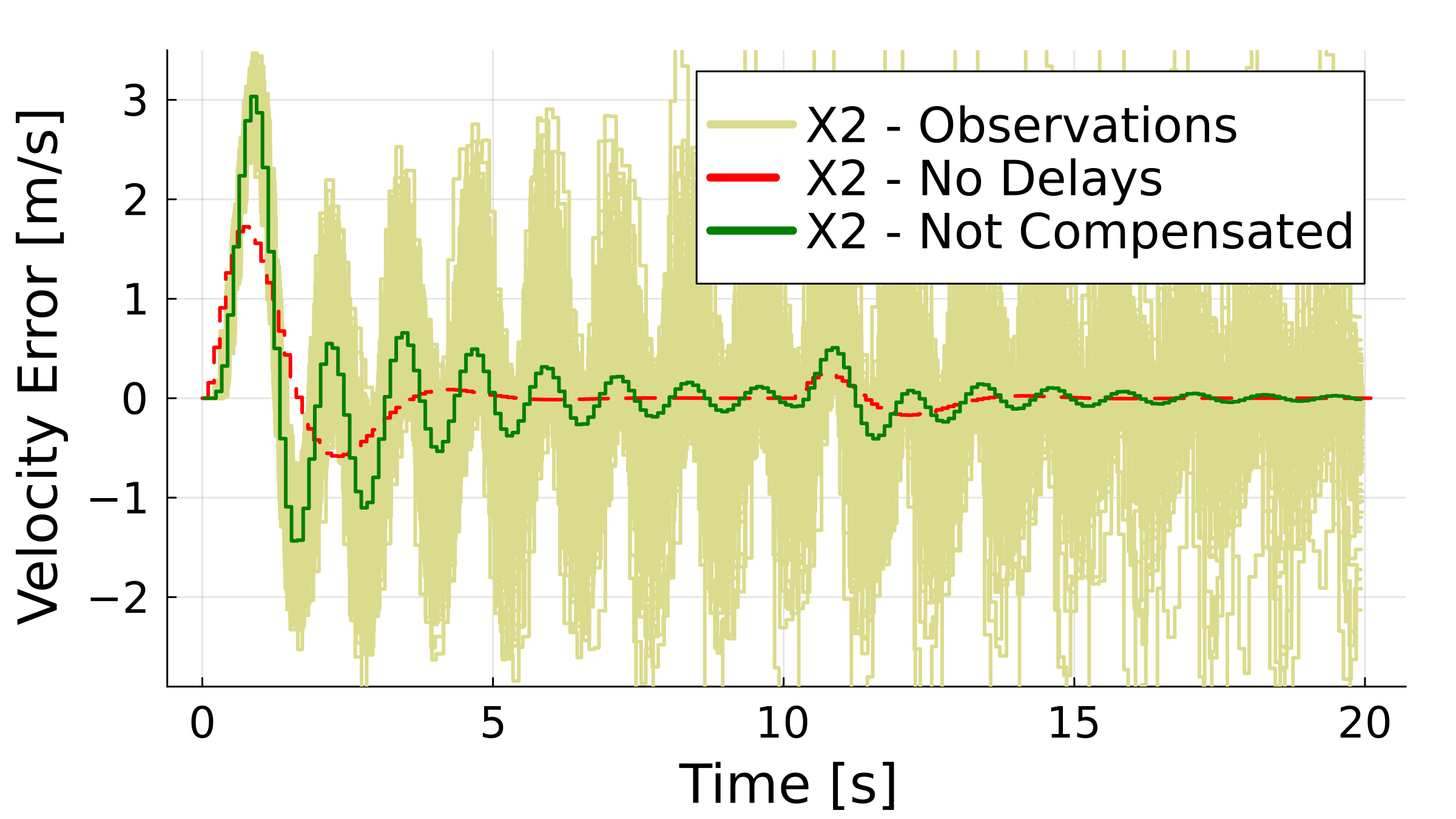}
    \label{fig:example:velocity-nc}}
    \subfloat[]{\includegraphics[width=0.335\linewidth]{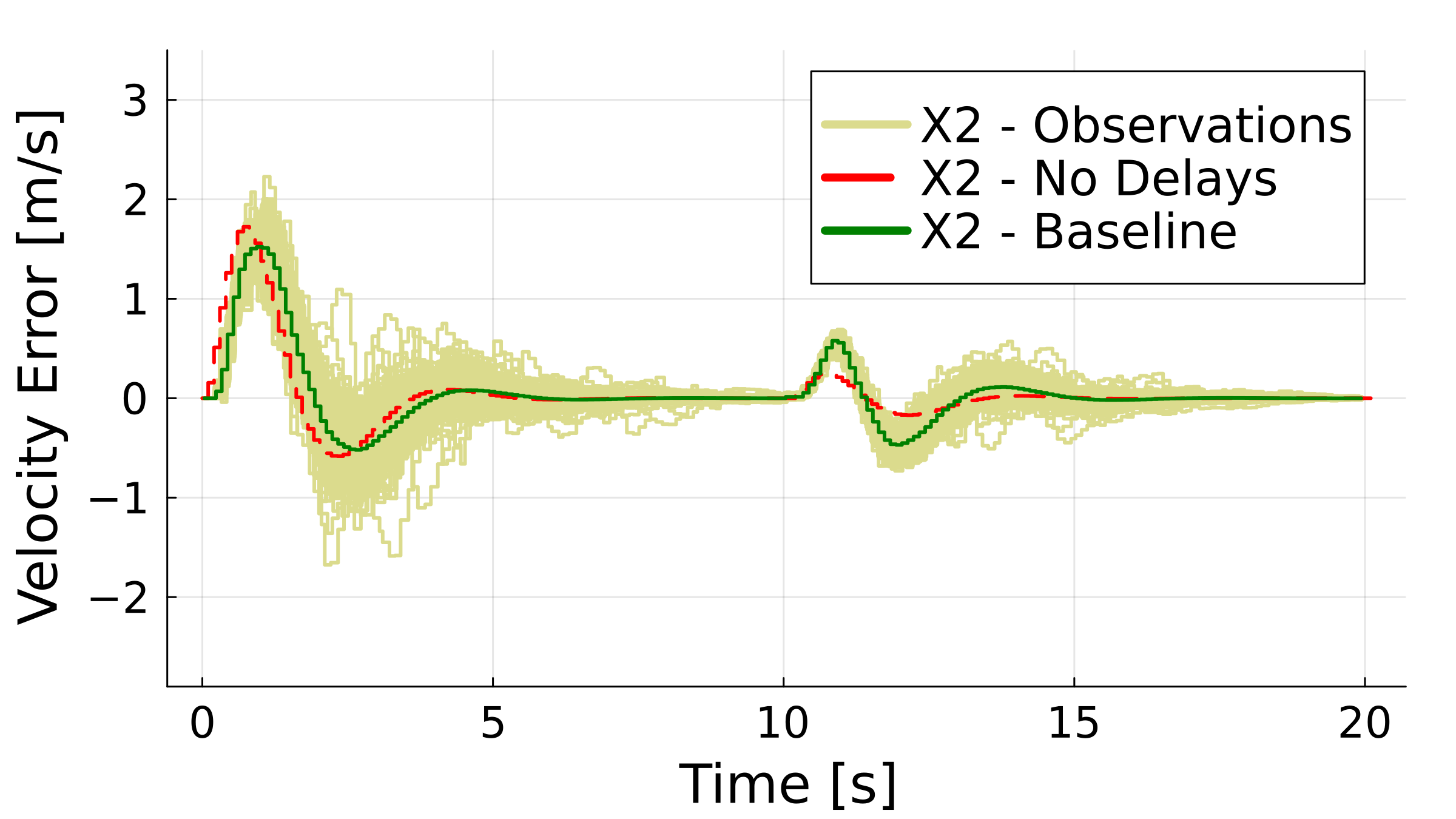}
    \label{fig:example:velocity-base}}
    \subfloat[]{\includegraphics[width=0.335\linewidth]{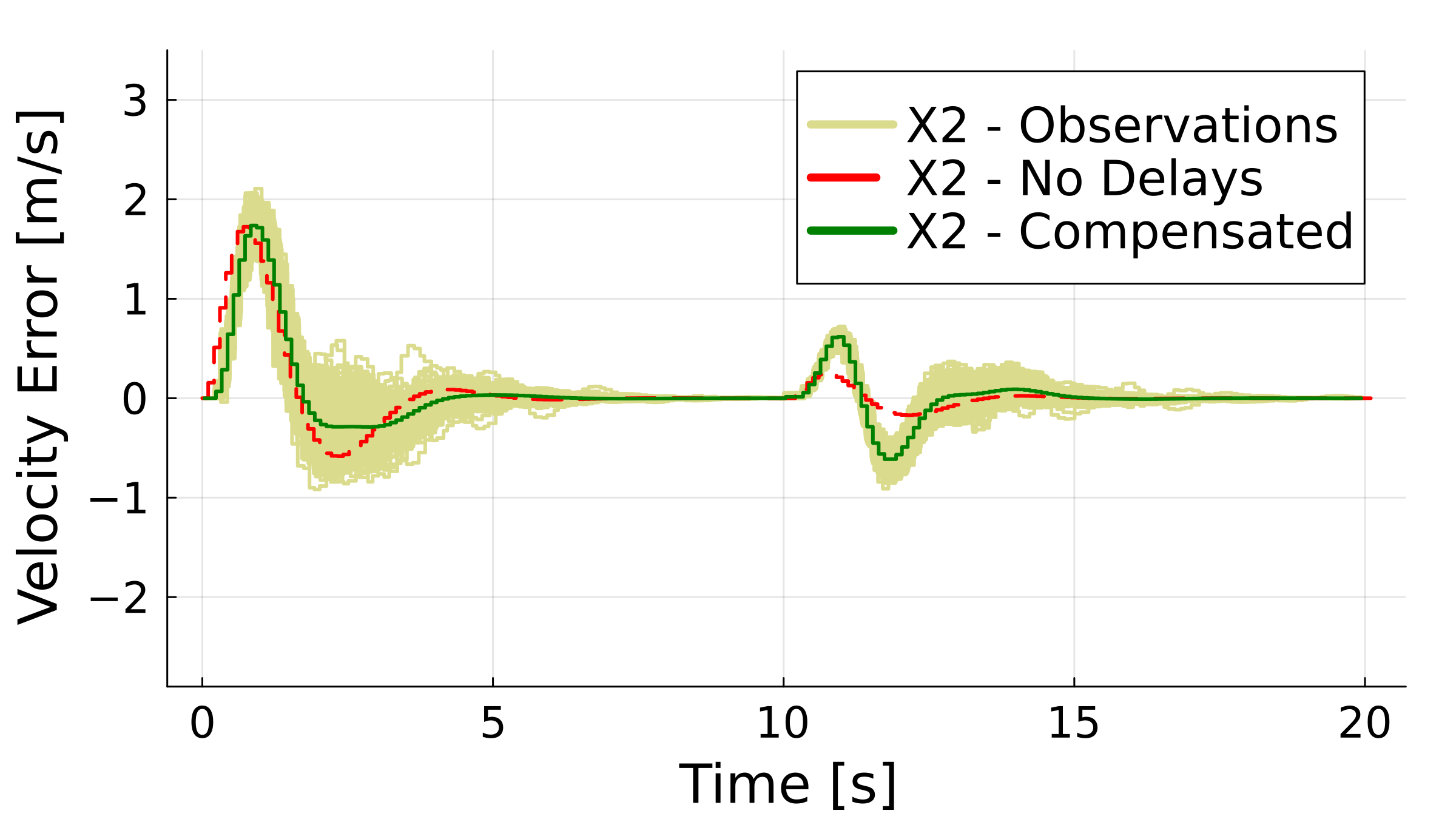}
    \label{fig:example:velocity-c}}

    \subfloat[]{\includegraphics[width=0.335\linewidth]{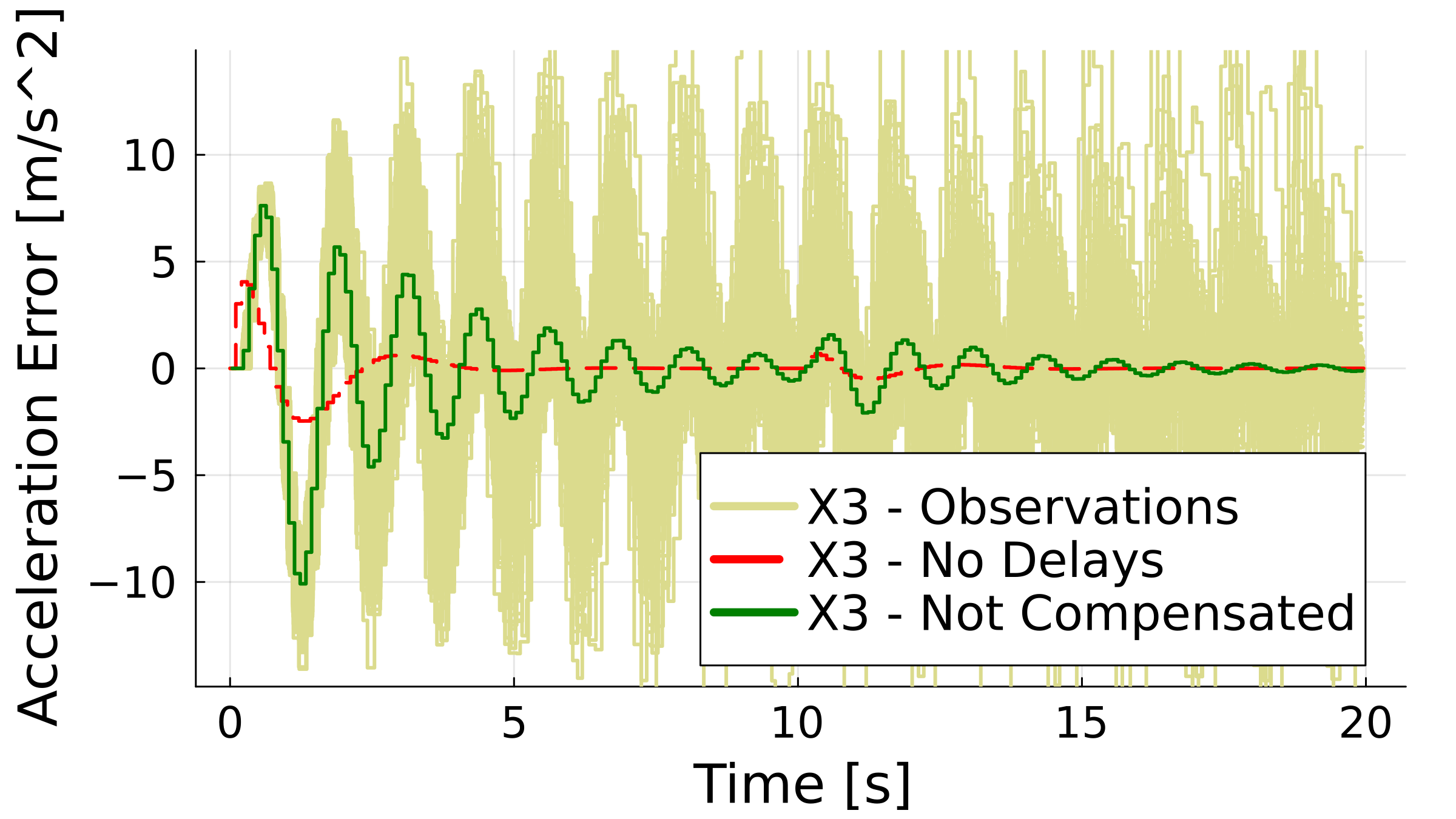}
    \label{fig:example:acceleration-nc}}
    \subfloat[]{\includegraphics[width=0.335\linewidth]{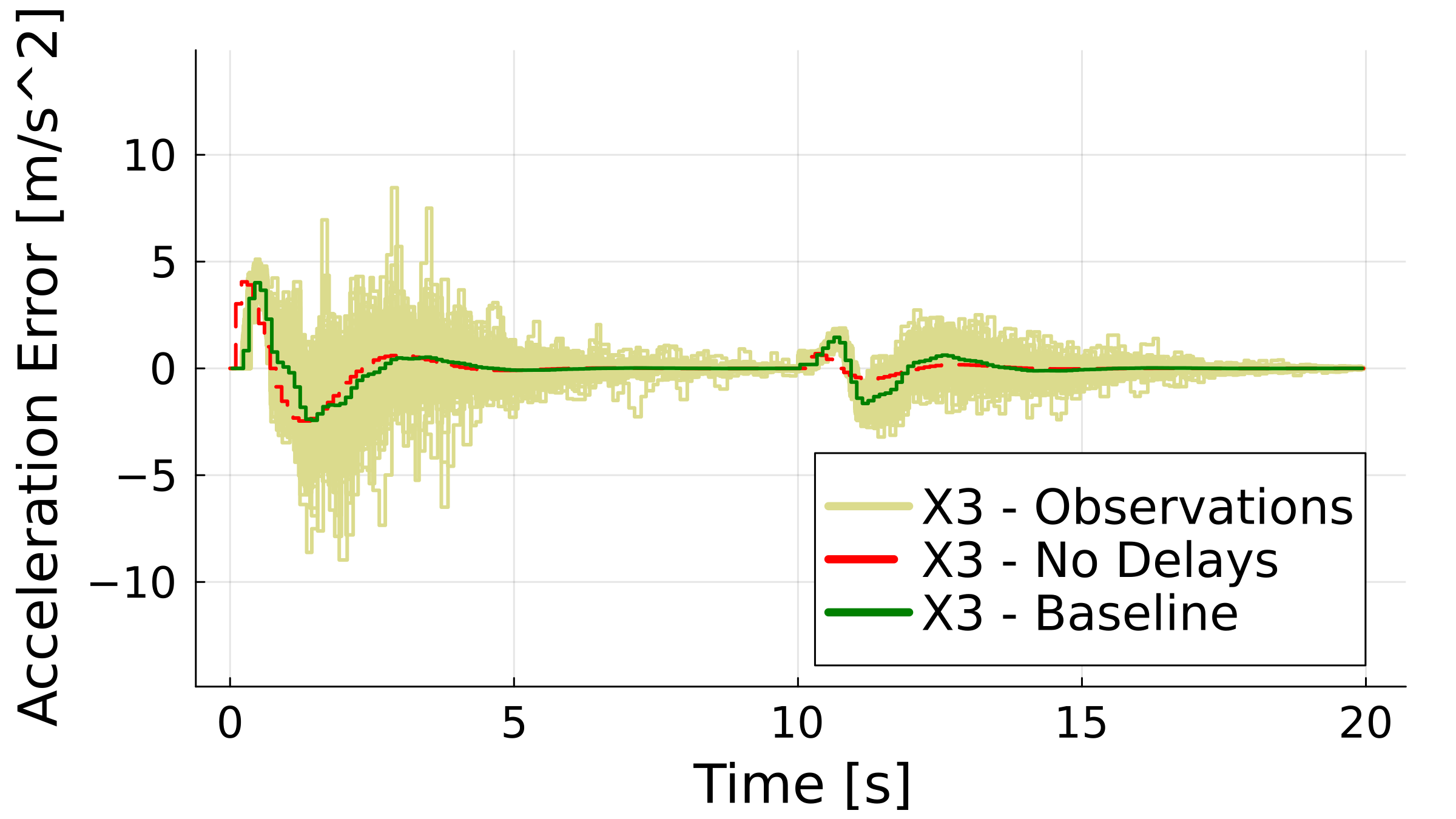}
    \label{fig:example:acceleration-base}}
    \subfloat[]{\includegraphics[width=0.335\linewidth]{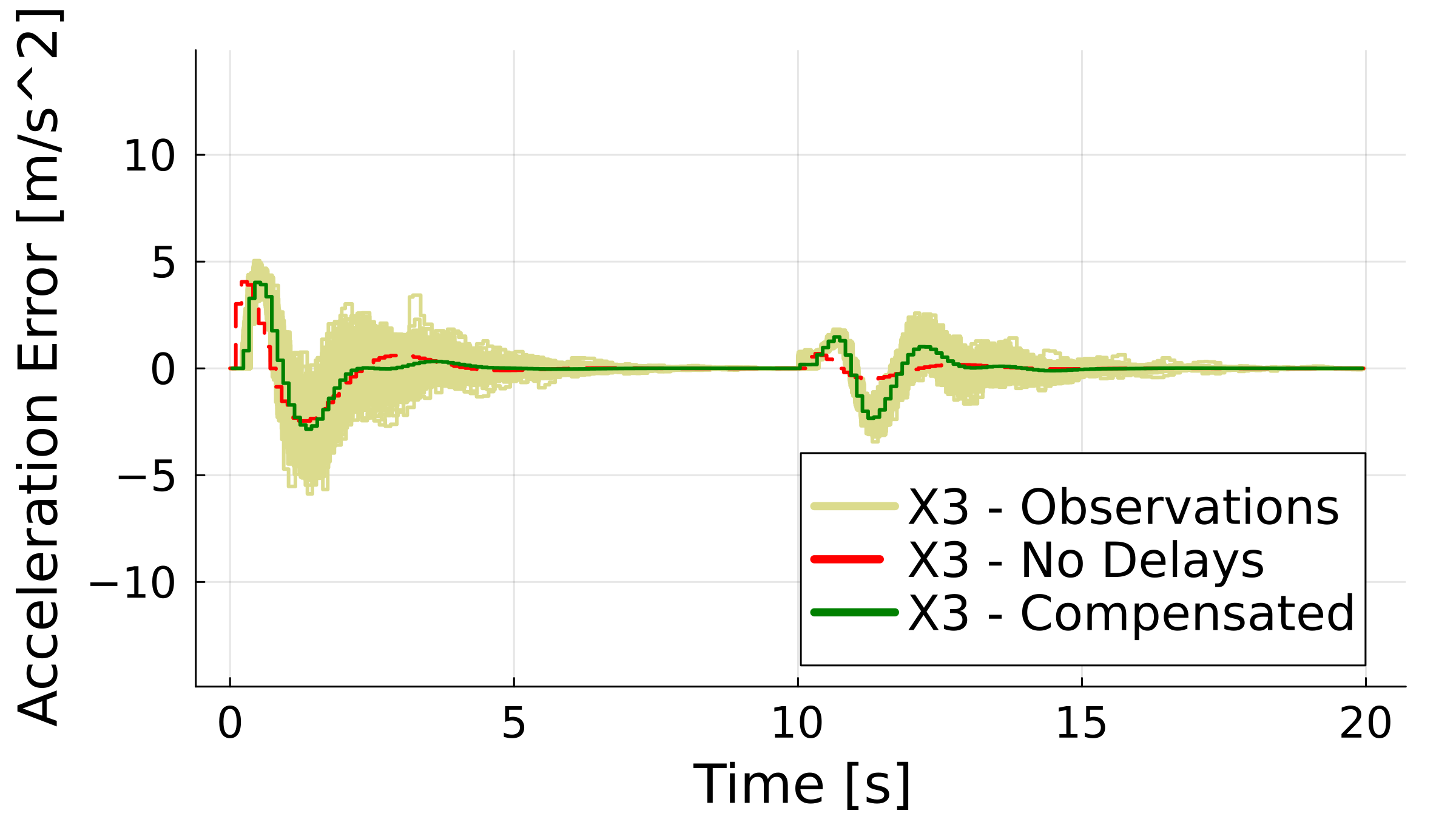}
    \label{fig:example:acceleration-c}}

\caption{Observations of the system states time response for the numerical examples. Inter-vehicle spacing ($x^1$) (a) in the absence of time delay compensation, (b) using the baseline time delay compensator and (c) using the \gls{3SFSP} time delay compensator.  Velocity error ($x^2$) (d) in the absence of time delay compensation, (e) using the baseline time delay compensator and (f) using the \gls{3SFSP} time delay compensator. Acceleration error ($x^3$) (g) in the absence of time delay compensation, (h) using the baseline time delay compensator and (i) using the \gls{3SFSP} time delay compensator.}
    \label{fig:example:states-response}
\end{figure*}

\begin{figure*}
    \centering

    \subfloat[]{\includegraphics[width=0.335\linewidth]{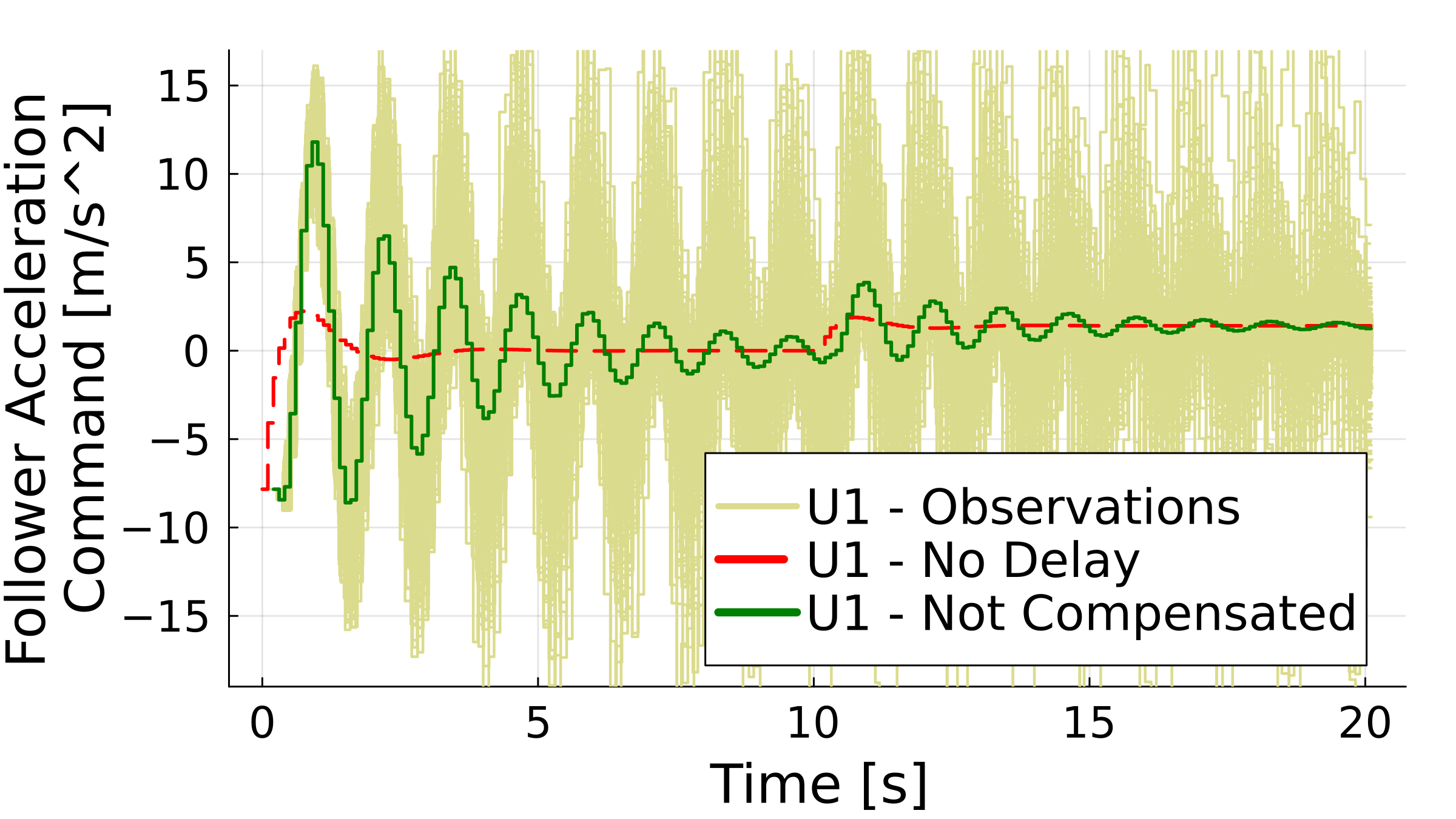}
    \label{fig:example:follower-control-nc}}
    \subfloat[]{\includegraphics[width=0.335\linewidth]{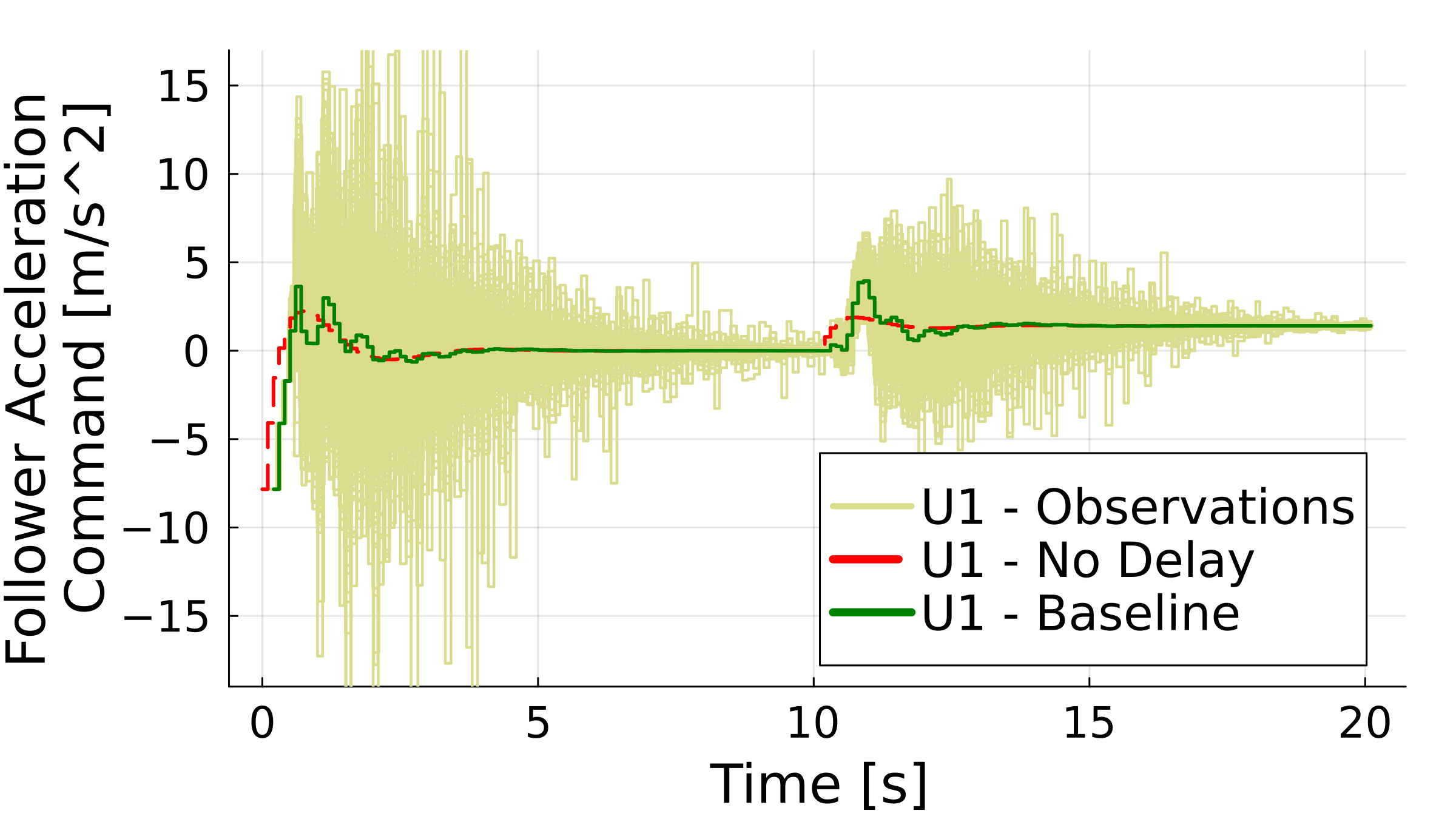}
    \label{fig:example:follower-control-base}}
    \subfloat[]{\includegraphics[width=0.335\linewidth]{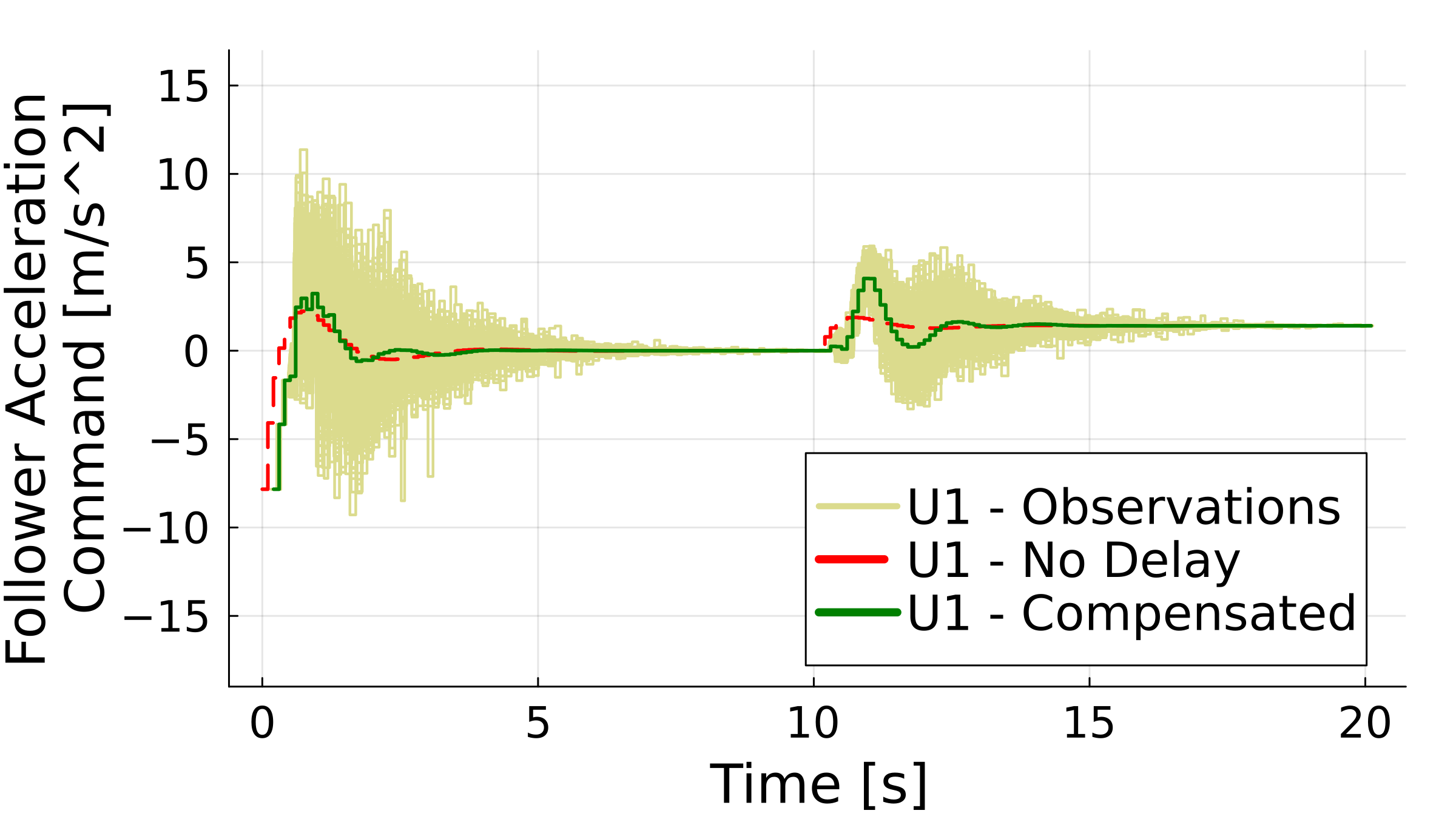}
    \label{fig:example:follower-control-c}}

    \subfloat[]{\includegraphics[width=0.335\linewidth]{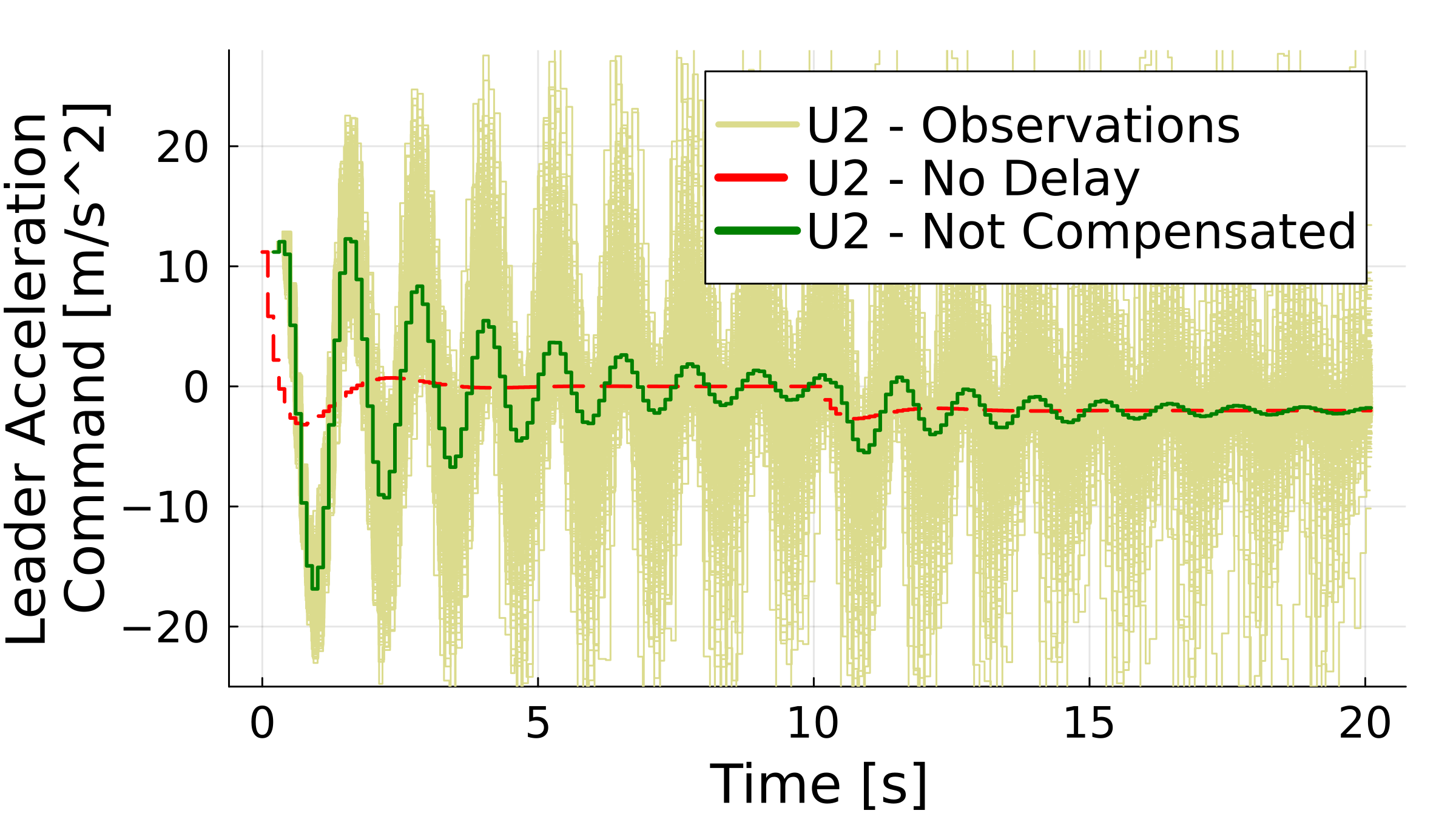}
    \label{fig:example:leader-control-nc}}
    \subfloat[]{\includegraphics[width=0.335\linewidth]{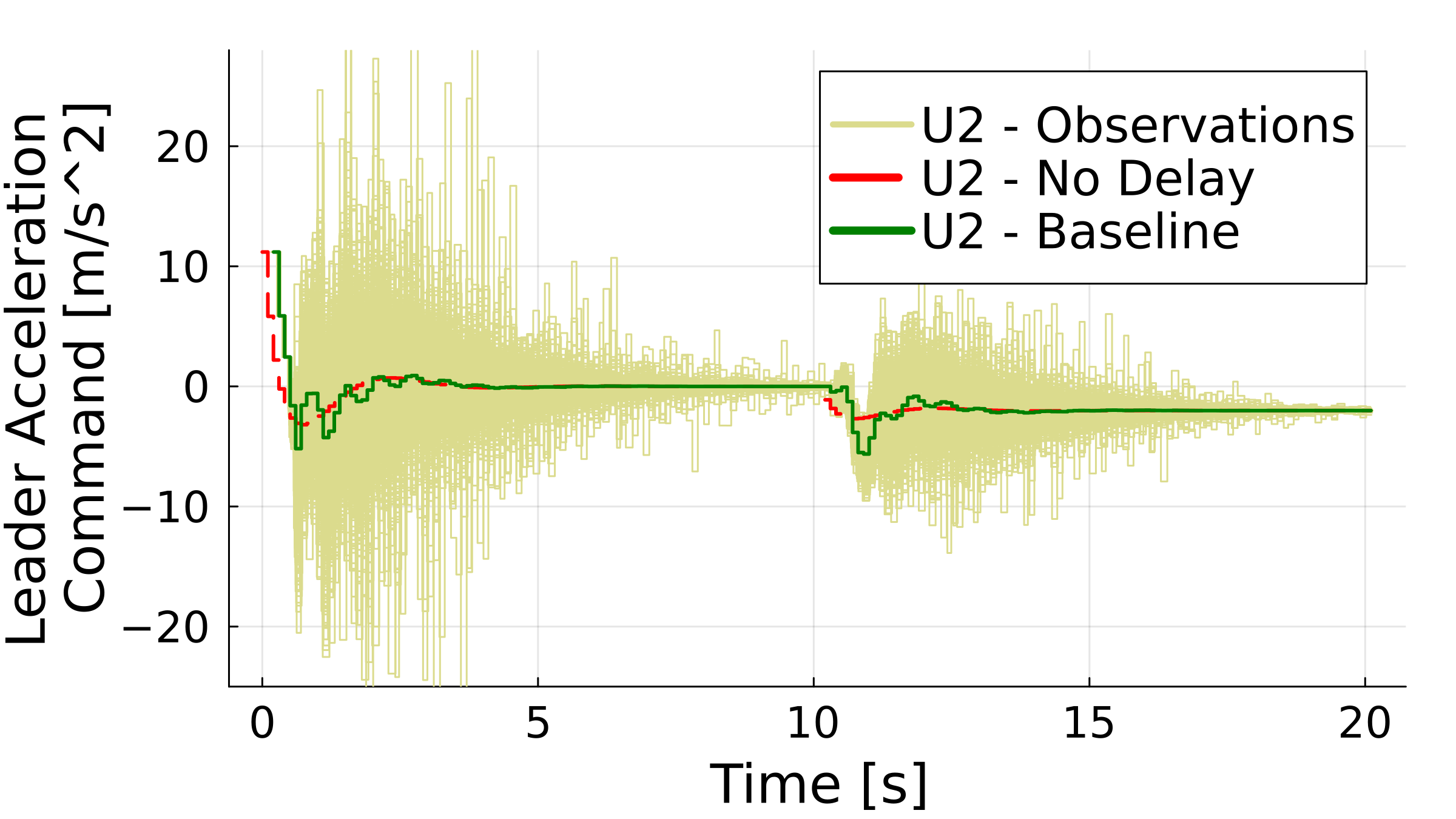}
    \label{fig:example:leader-control-base}}
    \subfloat[]{\includegraphics[width=0.335\linewidth]{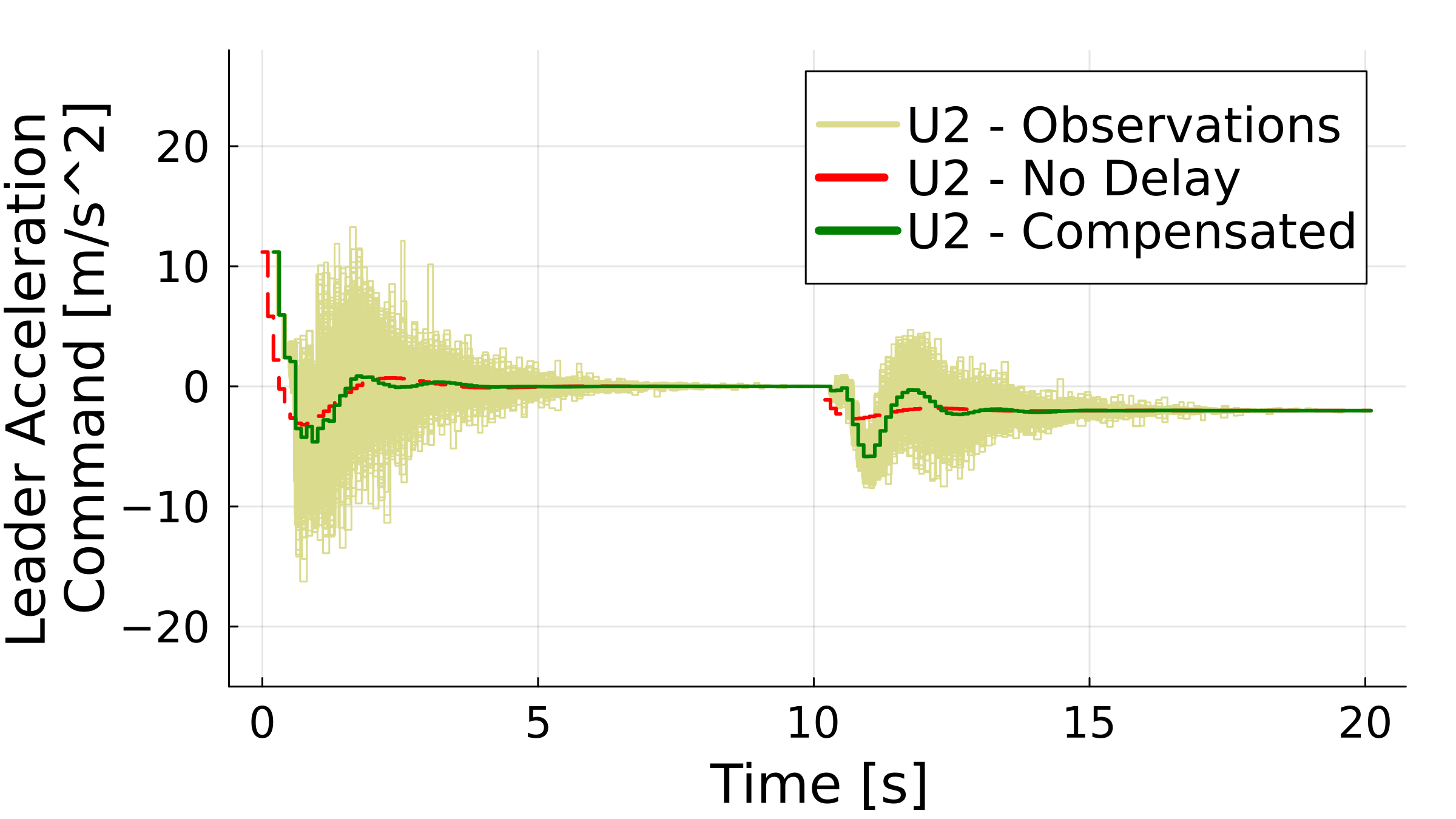}
    \label{fig:example:leader-control-c}}
    
        \caption{Observations of the control signal time response for the numerical examples. Follower vehicle acceleration command ($u^1$) (a) in the absence of time delay compensation, (b) using the baseline time delay compensator and (c) using the \gls{3SFSP} time delay compensator. Leader vehicle acceleration command ($u^2$) (d) in the absence of time delay compensation, (e) using the baseline time delay compensator and (f) using the \gls{3SFSP} time delay compensator.}
    \label{fig:example:inputs-response}
\end{figure*}

Then, to verify the effectiveness of the \gls{3SFSP}, a set of numerical simulations was performed to compare the response of the system in ideal conditions, i.e. constant sampling intervals and no time delays, with the response of the system with and without the compensation scheme provided by the \gls{3SFSP}. Furthermore, a comparison with a classical \gls{FSP}, referred to as the baseline compensator, considering a fixed delay of two sampling periods and a pole location for the filter at $\alpha=0.3$ was also performed. For each case, 5000 observations of the system response were evaluated, enough for the convergence of the ensemble average of observations. Accordingly, Fig.  \ref{fig:example:states-response} presents the response of the systems to a unit step in the reference inter-vehicle distance and to a constant disturbance of magnitude $10 m/s^2$ applied at the instant $t = 10$s. The corresponding control signals resulting from the simulation are presented in Fig.  \ref{fig:example:inputs-response}.

\begin{figure*}
\centering
\subfloat[]{\includegraphics[width=0.335\linewidth]{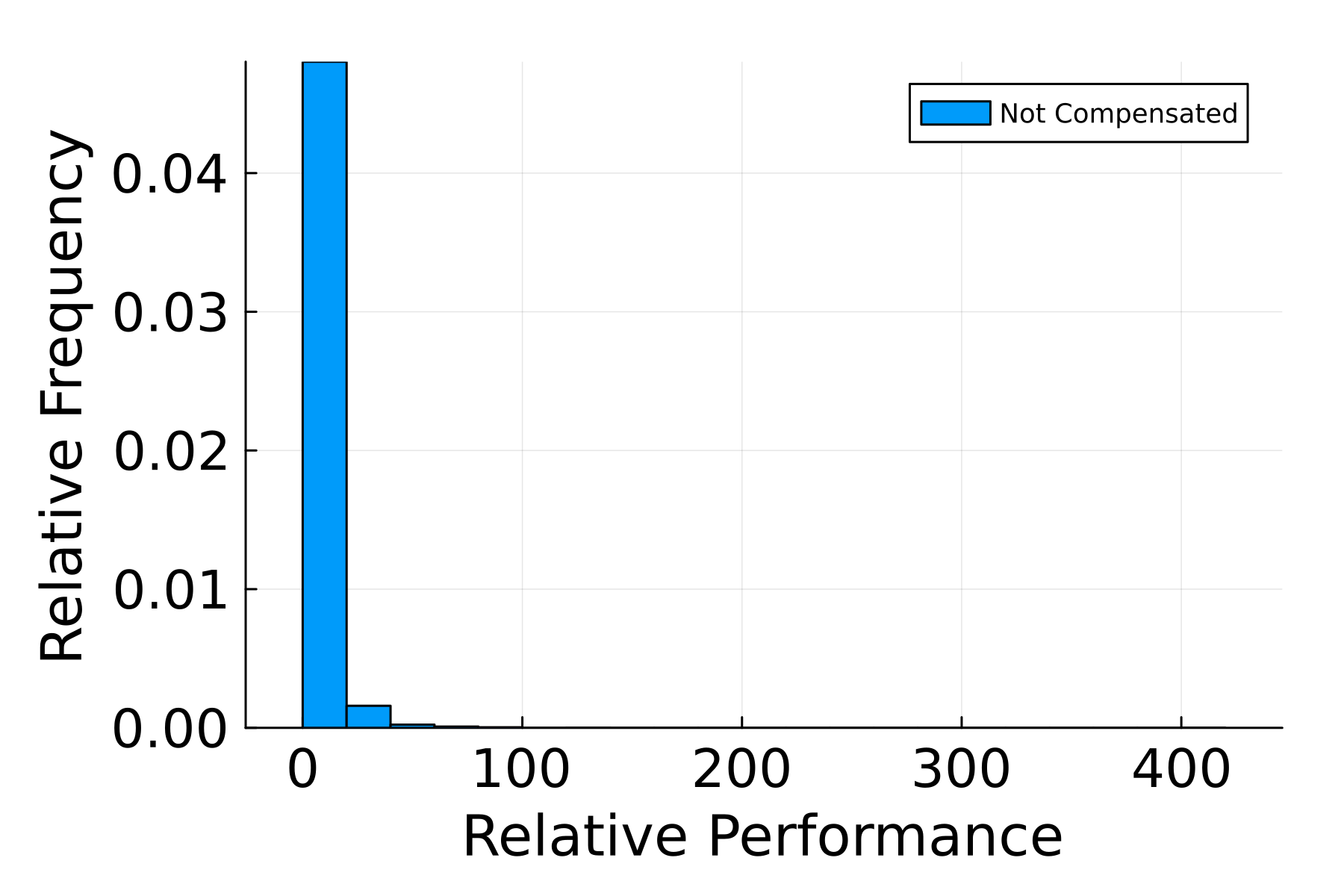}%
\label{fig:example:filter-parameteric-singular-value}}
\subfloat[]{\includegraphics[width=0.335\linewidth]{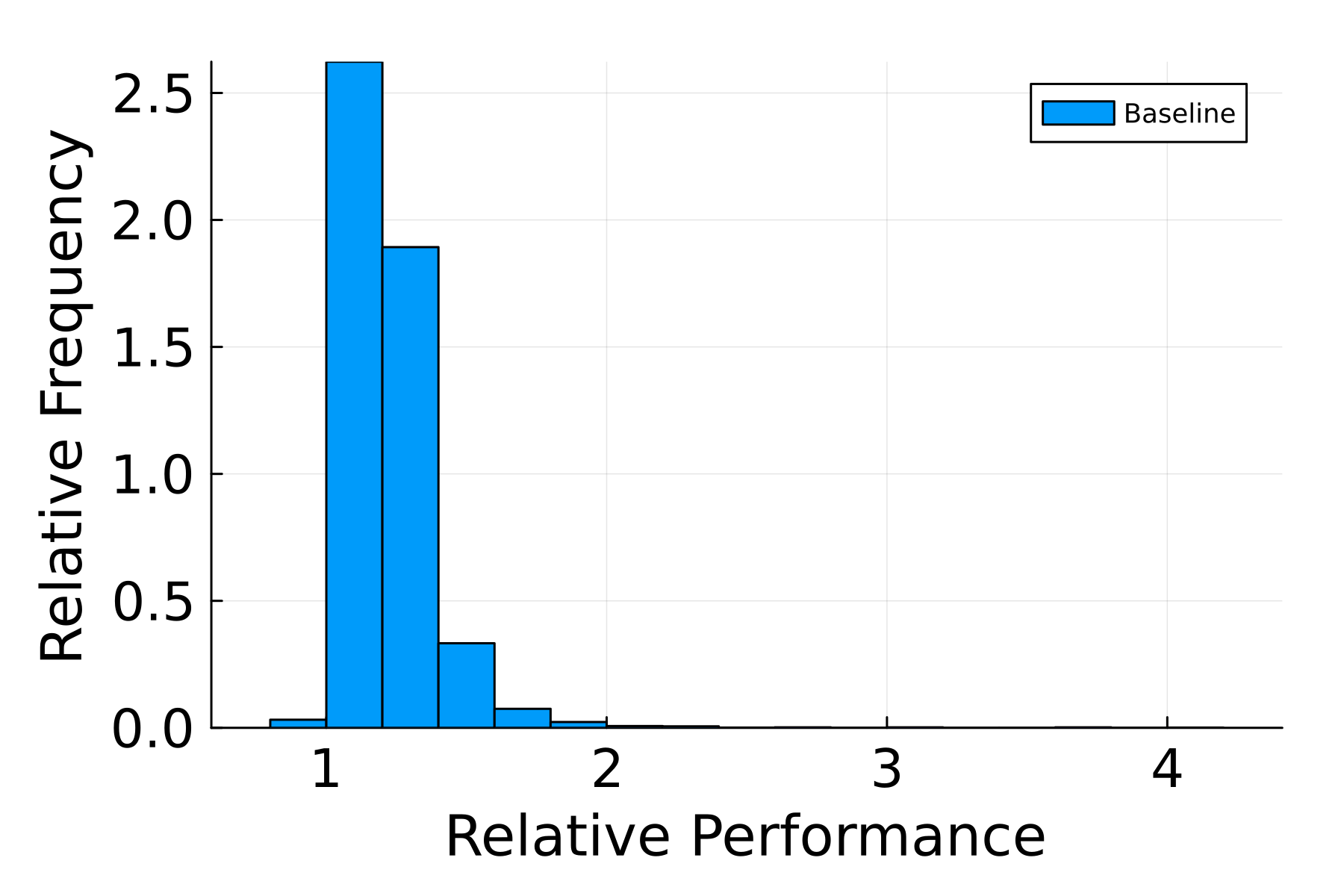}%
\label{fig:example:filter-parameteric-spectral-radius}}
\subfloat[]{\includegraphics[width=0.335\linewidth]{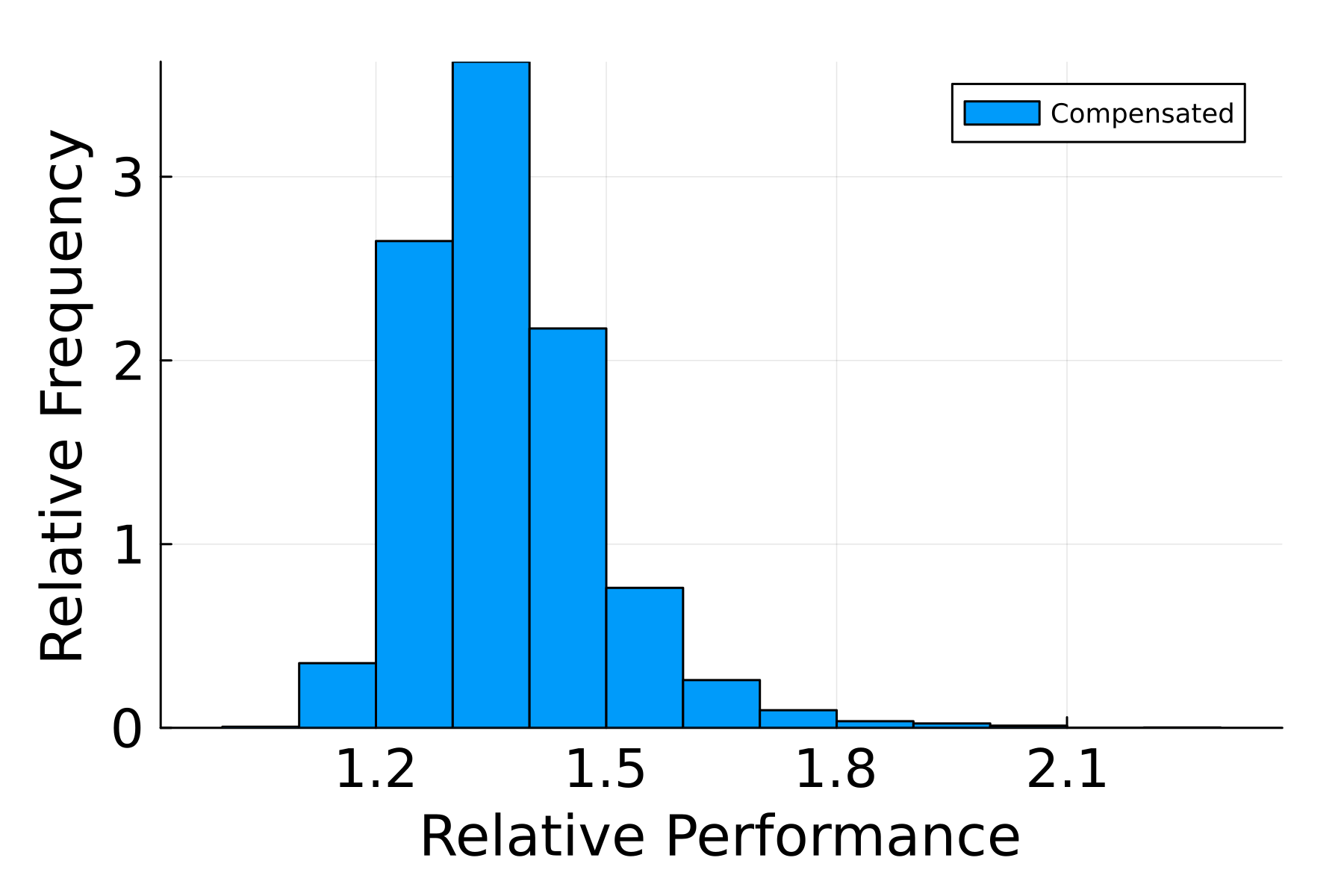}%
\label{fig:example:filter-parameteric-spectral-radius}}
\caption{Performance relative to the case with fixed sampling period and without time delays. (a) No compensation. (b) Baseline compensator. (c) \gls{3SFSP} compensator.}
\label{fig:example:performance-analysis}
\end{figure*}

\begin{figure}
    \centering
    \includegraphics[width=0.8\linewidth]{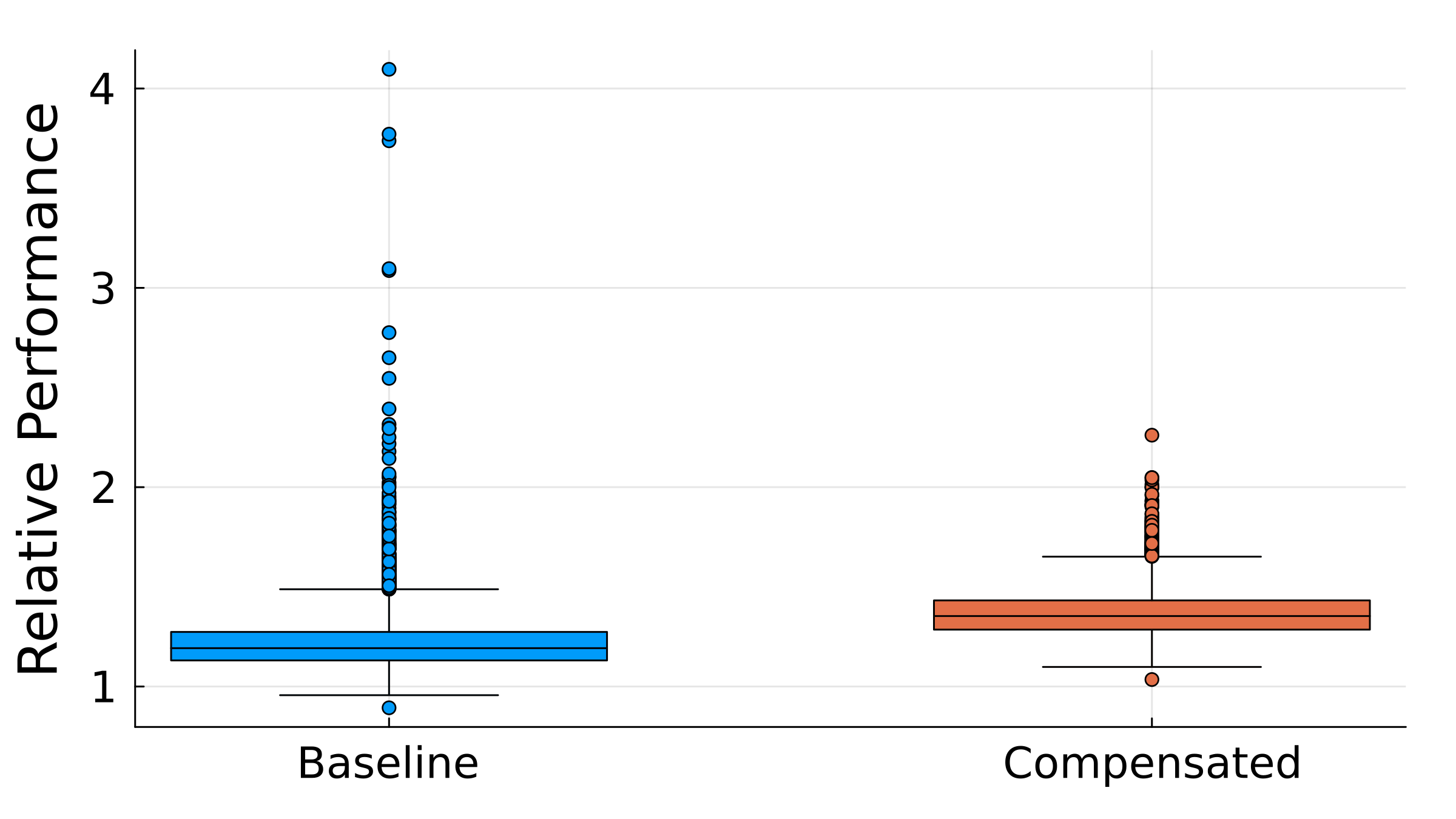}
    \caption{Box plot for the performance relative to the case with fixed sampling period and without time delays considering the baseline and \gls{3SFSP} compensators}
    \label{fig:example:performance-comparison}
\end{figure}

The results presented in Fig.  \ref{fig:example:states-response} highlight that the oscillatory behavior induced by the time delays in the absence of compensation is completely removed once the \gls{3SFSP} or the \gls{FSP} is used for compensation. Furthermore, the application of the \gls{3SFSP} reduced the overall variance observed across different observations of the system response. When compared to the classical \gls{FSP}, the response of the system compensated by \gls{3SFSP} presents a reduced variance across different observations of the system response.

Although the step response ensemble average of the compensated system in Fig.  \ref{fig:example:states-response} closely resembles the ideal step response, the response to disturbances does not. This occurs due to the fact that for the \gls{3SFSP}, like the \gls{FSP}, the response to disturbances is heavily influenced by the response of the prediction error filter whose zeros are constrained by the design requirement of avoiding the appearance of undesired poles of the plant model in the predictor dynamics. But even with the observed amplification of disturbances in the compensated scenario, the overall disturbance rejection performance of the system increases once the compensation scheme of the \gls{3SFSP} is employed.

For a quantitative evaluation of the system's performance, each observation of the system response was evaluated with respect to the metric presented in Eq. \eqref{example:performance-metric}, that measures the energy of the tracking error signal.

\begin{equation}\label{example:performance-metric}
    J = \sum_{n=0}^N \|\boldsymbol{B}^r r_k - \boldsymbol{x}\|^2.
\end{equation}

Fig. \ref{fig:example:performance-analysis} presents the performance of the system relative to the performance of the system in ideal conditions, denoted by $J_{\text{ideal}}$, computed as the ratio $J/J_{\text{ideal}}$, while Fig \ref{fig:example:performance-comparison} presents a performance comparison between the \gls{3SFSP} and \gls{FSP} (baseline). The quantitative performance comparison clearly states the performance improvement observed through the use of the \gls{3SFSP} or the \gls{FSP}.
For the case in which no compensation is employed, the average and worst tracking error signal energy observed are 9.27 and 407 times that of the system in ideal conditions, respectively. For the case in which the \gls{3SFSP} is employed, the average and worst tracking error signal energy observed are 1.37 and 2.26 times that of the system in ideal conditions, respectively. For the case in which the \gls{FSP} is employed, the average and worst tracking error signal energy observed are 1.23 and 4.10 times that of the system in ideal conditions, respectively. The quantitative analysis confirms the qualitative observation that the \gls{3SFSP} provides a smaller variability on the system performance when compared with the \gls{FSP}, demonstrating how, beyond effectively compensating stochastic time delays, the \gls{3SFSP} attenuates the uncertainty propagation from the time delays to the system response.

\section{Conclusion}\label{sec:conclusion}

This work presented modeling, design and analysis procedures for stochastic \gls{NCS} considering a compensation scheme based on the \gls{FSP}. The effectiveness of the proposed compensation scheme was evaluated through a numerical simulation considering a well-established model for a \gls{CACCS}. The presented results demonstrate that by employing the \gls{3SFSP} for time delay compensation, the performance of the closed-loop system can be made very close to that of an ideal realization of the system. 

It is important to highlight that the proposed approach is applicable to \gls{LTI} plants in which the full state of the system is measurable and to \gls{NCS} with a very high degree of network reliability, in terms of how often messages fail to be delivered. Hence, further work is still necessary to broaden the range of systems that can benefit from the \gls{3SFSP} topology. Furthermore, in further work, we aim to relax the specifications imposed on the response times of the control system tasks by compensating violations of such specifications and investigate the representativeness of the \gls{iid} model for response time.

\section*{Acknowledgments}

\noindent The Authors acknowledge the financial support from \emph{Coordena\c{c}\~ao de Aperfei\c{c}oamento de Pessoal de N\'ivel Superior} – Brasil (CAPES) [finance code $001$ and grant $88881.878833/2023-01$ - OPCOMPLEX, SticAmSud], from the National Council for Scientific and Technological Development (CNPq, Brazil) [grants $304032/2019$-$0$, $403949/2021$-$1$, and $406477/2022$-$1$], and from FAPESC [project $2023TR001506$].
\begin{center}
\includegraphics[width=0.7\linewidth]{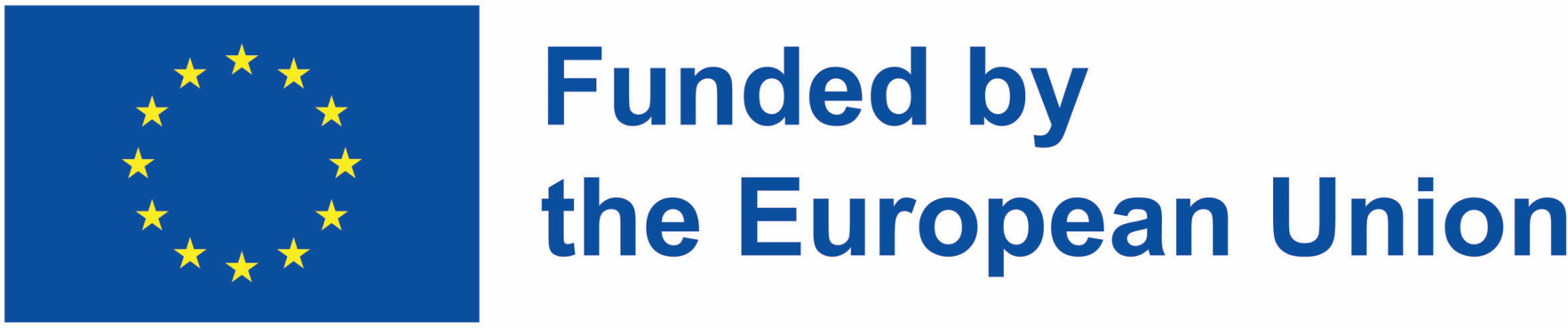}
\end{center}
\noindent This work has been funded by European Union’s Horizon Europe research and innovation programme under the \textit{Marie Sk\l{}odowska-Curie Action} grant agreement No $101149263$ (\textit{REAL OPT $4$ CONTROL}).

\bibliography{references} 
\bibliographystyle{IEEEtran}

\end{document}